\documentclass[aps,pre,preprint,showpacs,showkeys,floatfix]{revtex4-1}
\usepackage{graphicx}
\usepackage{dcolumn}
\usepackage{bm}
\usepackage{float}
\usepackage{color}
\usepackage{amssymb}
\usepackage{amsmath}
\usepackage{color}
\usepackage[caption=false]{subfig} 
\usepackage{xcolor}
\usepackage[percent]{overpic}
\usepackage{enumitem}
\usepackage{natbib}
\usepackage{overpic}
\usepackage[none]{hyphenat}
\newcommand{\valmed}[1]{\langle #1 \rangle}

\begin{document} 

\title{Disease spreading with social distancing: A prevention strategy in
  disordered multiplex networks}

\author{Ignacio A. Perez} \email{ignacioperez@mdp.edu.ar}
\affiliation{Instituto de Investigaciones F\'isicas de Mar del Plata
  (IFIMAR)-Departamento de F\'isica, FCEyN, Universidad Nacional de
  Mar del Plata-CONICET, D\'ean Funes 3350, (7600) Mar del Plata, Argentina.}

\author{Mat\'ias A. Di Muro} \affiliation{Instituto de
  Investigaciones F\'isicas de Mar del Plata (IFIMAR)-Departamento de
  F\'isica, FCEyN, Universidad Nacional de Mar del Plata-CONICET,
  D\'ean Funes 3350, (7600) Mar del Plata, Argentina}

\author{Cristian E. La Rocca} \affiliation{Instituto de
  Investigaciones F\'isicas de Mar del Plata (IFIMAR)-Departamento de
  F\'isica, FCEyN, Universidad Nacional de Mar del Plata-CONICET,
  D\'ean Funes 3350, (7600) Mar del Plata, Argentina}

\author{Lidia A. Braunstein} \affiliation{Instituto de Investigaciones
  F\'isicas de Mar del Plata (IFIMAR)-Departamento de F\'isica, FCEyN,
  Universidad Nacional de Mar del Plata-CONICET, D\'ean Funes 3350,
  (7600) Mar del Plata, Argentina}  \affiliation{Physics Department,
  Boston University, 590 Commonwealth Ave., Boston, Massachussets
  02215, USA}

\begin{abstract}
  
\noindent

The frequent emergence of diseases with the potential to become
threats at local and global scales, such as influenza A(H1N1), SARS,
MERS, and recently COVID-19 disease, makes it crucial to keep
designing models of disease propagation and strategies to prevent or
mitigate their effects in populations. Since isolated systems are
exceptionally rare to find in any context, especially in human contact
networks, here we examine the susceptible-infected-recovered model of
disease spreading in a \emph{multiplex network} formed by two distinct
networks or layers, interconnected through a fraction $q$ of shared
individuals (overlap). We model the interactions through weighted
networks, because person-to-person interactions are diverse (or
\emph{disordered}); weights represent the contact times of the
interactions. Using branching theory supported by simulations, we
analyze a social distancing strategy that reduces the average contact
time in both layers, where the intensity of the distancing is related
to the topology of the layers. We find that the critical values of the
distancing intensities, above which an epidemic can be prevented,
increase with the overlap $q$. Also we study the effect of the social
distancing on the mutual giant component of susceptible individuals,
which is crucial to keep the functionality of the system. In addition,
we find that for relatively small values of the overlap $q$, social
distancing policies might not be needed at all to maintain the
functionality of the system.

\end{abstract}

\maketitle

\section{Introduction}
\label{intro}

Localized outbreaks of resurgent or new diseases often have the
potential to become epidemics, affecting a relevant proportion of a
given region, if no preventive measures are undertaken or mitigation
strategies are implemented. Worse still, the high interconnection
between cities and also between countries extremely favors the
spreading of a disease throughout the entire world, which may turn an
outbreak into a pandemic in a matter of months, weeks, or even
days. This is what recently happened with COVID-19 disease, declared a
pandemic on March 11, 2020, by the World Health
Organization~\cite{WHO-20,Cuc-20}. Over the last decades researchers
across multiple disciplines have been modeling disease propagation to
develop strategies that could prevent or at least curtail epidemics
(and pandemics, in the worst cases). Infectious diseases usually
propagate through physical contacts among
individuals~\cite{Bai-75,Ander-92}, and researchers have found that
modeling these contact patterns~\cite{Gon-08,Catt-10} is best achieved
using complex networks~\cite{Bocc-06,Bar-04,New-10,Pas-15}, in which
individuals and their interactions are represented by nodes and links,
respectively. Numerous disease propagation models have made use of
complex networks, including the susceptible-infected-susceptible
(SIS)~\cite{Bai-75,Bocc-06,Pas-15} and susceptible-infected-recovered
(SIR)~\cite{Ker-27,Gra-83,Ander-92,New-02,Bocc-06,Pas-15} models. In
these epidemic models, individuals can be in different states. For
example, \emph{infected} (I) individuals carry the disease and can
transmit it to \emph{susceptible} (S) neighbors that are not immune to
the disease, while \emph{recovered} (R) individuals do not participate
in the propagation process because they have either recovered from a
previous infection or because they have died. The SIR model, in which
individuals acquire permanent immunity after recovering from an
illness, is the simplest and most used to study non-recurrent
diseases. In the discrete-time version of this model~\cite{Bai-75}, at
each time step I individuals spread the disease to their S neighbors
with the same probability $\beta \in [0,1]$, and switch to the R state
$t_r$ time steps after being infected, where $t_r$ is the recovery
time of the disease. The propagation reaches the final stage when the
number of I individuals goes to zero. At this stage, the fraction $R$
of recovered individuals indicates the extent of the infection, since
all recovered individuals were once infected. In this model, the
spreading is controlled by the effective probability of infection $T =
1 - (1 - \beta)^{t_r}$, the \emph{transmissibility}, with $T \in
[0,1]$. When $T$ is below a critical value $T_c$, also called the
\emph{epidemic threshold}, the fraction $R$ of recovered individuals,
which is the order parameter of a continuous phase transition, is
negligible compared to the system size $N$, and therefore the system
is in a non-epidemic phase. On the other hand, above $T_c$ the
fraction $R$ is comparable to $N$ and thus it is said that the disease
becomes an epidemic~\cite{New-02,Mil-07,Ken-07,Lag-09}.

In countries such as Italy or Spain, COVID-19 disease spread
uncontrollably, causing tens of thousands of deaths, not only because
of the intrinsic virulence of the disease, but also due to the
collapse of the health system. On the other hand, in countries where
the disease arrived after spreading over Europe and the United States,
such as Argentina, authorities immediately implemented a preventive
massive lockdown, limiting the contact between people. The goal was to
avoid the collapse of the health system, to be able to provide rapid
and effective medical response to those affected by the
disease. Nonetheless, due to the vulnerable economic and social
conditions of a vast portion of the Argentinian society, it was
practically impossible to completely cut off the interactions between
people, which enabled the disease to keep disseminating through the
population. Therefore, not only is it important to understand how a
disease spreads under different isolation conditions, but it is also
crucial to evaluate how it can be
mitigated. Vaccination~\cite{Coh-03,Fer-06,Ban-06,Pas-15,Buo-15,Zuz-15,Mer-16,DiMu-18,Zuz-19}
is regarded as the most efficient measure to prevent or attenuate an
epidemic, providing individuals with immunity against a disease. In
addition, this pharmacological intervention avoids the negative
consequences that society may face after implementing strict and
detrimental policies such as partial or complete lockdown and
quarantine (e.g., social and economic disruption). However, in
emergency situations such as the ongoing COVID-19 pandemic, a vaccine
is not yet available and therefore, it is necessary to take other
types of countermeasures.

On the other hand, there is a group of strategies less severe than
quarantine and lockdown, but in the same spirit. ``Social distancing'' 
strategies~\cite{Gro-06,Eas-10,Lag-11,Buo-12,Val-12,Buo-13,Per-19}
are a set of actions intended to reduce contact between individuals
(shorten the interaction times, maintain a minimum physical distance,
avoid crowded places, etc.), with the aim of decreasing the
probability of disease transmission. These kinds of strategies respond
to the direct transmission mechanism of the virus that causes the
COVID-19 disease, although they also apply to diseases that spread in
a similar way. The SARS-CoV-2 virus is expelled in the form of
droplets from an infected individual through mouth and nose (when
talking, exhaling, or coughing), and can enter a healthy individual
(located within a 1-2 m range, approximately, and facing the
infected individual) through the mucous membranes (eyes, nose, and
mouth)~\cite{WHO-20-bis}. Therefore, as individuals are more deeply
interconnected, the probability of infection significantly
increases. Social distancing, along with partial or complete lockdown,
has been undertaken in many countries to face the COVID-19
pandemic. Researchers believe that, until a vaccine is widely
available, social distancing will remain one of the primary
measures to combat the spread of the disease~\cite{Badr-20}.

One way of modeling social distancing measures is by reducing contact
times between individuals. In real-world networks, contact times
usually span a broad
distribution~\cite{Catt-10,Kars-11,Steh-11}. These kinds of systems are
known as \emph{disordered} networks, which are characterized by the
existing diversity in the strength or intensity of the interactions
among the different parts of the system. Disordered networks have
been receiving much attention
recently~\cite{Bar-04,Buo-12,Buo-13,Per-19}. In Ref.~\cite{Per-19}
disorder is implemented using a weighted complex
network~\cite{Bar-04}, in which the weights associated with links
represent the normalized contact times $\omega$ of the interaction
between two individuals. These contact times follow a power-law
distribution, $W(\omega) = 1/(a \omega)$ that resembles experimental
results~\cite{Catt-10,Kars-11,Steh-11}. The parameter $a$ is the
\emph{disorder intensity}, which controls the range of contact times
in the distribution along with its average value. Also, the
interactions between individuals are categorized as either
\emph{close} (larger average contact time) or \emph{distant} (shorter
average contact time), each representing complementary fractions
($f_1$ and $1 - f_1$, respectively) of the total number of
interactions. This is carried out by controlling the corresponding
disorder intensity of the contact time distributions. Researchers
have found that for a system in an epidemic phase, when the fraction
$f_1$ of close contacts is sufficiently small, increasing the disorder
intensity of the distant interactions to decrease their average
contact time may switch the system to a non-epidemic phase.

A significantly relevant magnitude to also consider is the size GCS
of the giant component of susceptible individuals, or largest
connected cluster, at the final stage. This cluster is formed by all
the remaining susceptible (healthy) individuals that are connected
with each other, and it is the network that sustains the functionality
of a society, e.g., the economy of a region. Using a generating
function formalism, Newman~\cite{New-05} showed that in the SIR model
there exists a second threshold $T^*$ above which the giant cluster of
susceptible individuals vanishes at the final stage. On the other
hand, Valdez \emph{et al.}~\cite{Val-12} showed that $T^*$ is an
important parameter to determine the efficiency of a mitigation or
control strategy, because any strategy that manages to decrease the
transmissibility below $T^*$ can protect a large and connected cluster
of susceptible individuals, even when the system is in an epidemic
state.

The previously mentioned studies were carried out using isolated
networks, i.e., networks that do not interact with other different
networks. Researchers have noted that isolated network models ignore
the ``external'' connections that real-world systems use to
communicate with their environment, which usually affect the behavior
of the dynamical processes that take place on complex
systems~\cite{Jia-14,Kiv-14,Bocc-14,Kene-15}. Thus, the modeling of
interconnected networks, i.e., \emph{networks of networks} (NoN) or
\emph{multilayer networks}, has become extremely relevant as it allows
a more accurate representation of real systems. The ubiquity of the
NoN has encouraged researchers to use them in the study of several
topics such as cascading failure~\cite{Bul-10,Bru-12,DiMu-16}, social
dynamics~\cite{Cast-09,Gal-08}, and disease
propagation~\cite{Sau-12,Coz-13,Are-16}. In particular, the SIR model
was simulated and solved theoretically in an overlapped
\emph{multiplex network}~\cite{Buo-14} system consisting of two
individual networks or layers, in which a fraction $q$ of
\emph{shared} nodes ($q$ overlap) is present in both layers. These
shared individuals connect the different layers, and their presence
makes diseases more likely to spread as $q$ increases~\cite{Buo-14}.

We believe that the aforementioned aspects are rather relevant to
consider, especially taking into account the ongoing COVID-19
pandemic. We aim to include them in a model that reflects, to some
extent, the situation that many regions are facing nowadays. In this
paper we study a disease-spreading process using the SIR model in an
overlapped two-layer multiplex network. The layers, $A$ and $B$, have
definite degree distributions and are connected through a fraction $q$
of shared nodes. They also have a distribution of normalized contact
times, with disorder intensities $a_A$ and $a_B$, which are related
through the topology of both layers. Based on this model, we implement
a social distancing strategy in which the average contact times
between individuals in both layers are reduced, and the disorder
intensities may be different in each layer. This could reflect the
fact that people within their homes have more prolonged interactions,
while protective measures increase to reduce the contact times when
they go to work or they use public transport. We use the branching
theory, supported by extensive simulations, to study this social
distancing strategy, and evaluate its effectiveness in preventing the
onset of an epidemic. In addition, the ultimate goal is to analyze
how the social distancing policies can protect the mutual giant
component of susceptible individuals, formed by the healthy
individuals participating in both layers, which is what keeps the
economy of a region running. We apply this strategy in a variety of
multiplex networks with different degree distributions, to explore the
effect of network structure.

\section{Model: Social distancing strategy}
\label{model}

We use an overlapped multiplex network formed by two layers, $A$ and
$B$, in which a fraction $q$ of nodes, called shared nodes, are
present in both layers. For the construction of the layers, both of
which are of size $N$, we use the Molloy-Reed
algorithm~\cite{Moll-95}. Each layer has its own uncorrelated degree
distribution $P_i(k)$, which gives the probability that a node has $k$
neighbors in layer $i = A, B$. The connectivity $k$ is limited by a
minimum and a maximum value, which we denote as $k^i_{\rm min}$ and
$k^i_{\rm max}$ respectively. On the one hand, we use a homogeneous
Poisson distribution $P(k) = e^{-\valmed{k}} \valmed{k}^k / k!$, in
which the average connectivity $\valmed{k}$ is also the most likely
value. A network with such degree distribution is known as an ER
(Erd\H os-R\'enyi) network~\cite{Erd-59}. In addition, we use a
heterogeneous power-law or scale-free (SF) distribution with
exponential cutoff $k_c$, $P(k) \propto k^{- \lambda} e^{- k / k_c}$,
which is more representative of real-world networks as usually some
individuals may have a high number of contacts, while the majority may
have only a few~\cite{New-02}. We consider the exponential cutoff
$k_c$ in the power-law distribution as real-world networks are finite
and the maximum number of connections of a node is limited by the size
of the system. Additionally, each layer is a weighted network in which
a link $j$ of layer $i$ has associated a weight $\omega_{ij}$, where
$\omega_{ij}$ is a normalized contact time that defines the intensity
of the interaction between the two individuals connected by that
link. The $\omega_{ij}$ values are taken from the theoretical
distribution $W_i(\omega_i) = 1/(a_i
\,\omega_i)$~\cite{Buo-12,Per-19}, $\omega_i \, \epsilon \,
        [e^{-a_i},1]$, where $a_i$ is the disorder intensity of layer
        $i = A, B$. Therefore, each link of the network is assigned a
        weight $\omega_{ij} = e^{-a_ir_j}$~\cite{Brau-07}, where $r_j$
        is a random number within the interval $[0,1]$. 

To simulate the disease spread, we use the SIR model taking into
account that the probability of infection depends not only on the type
of disease, but also on the intensity of the interactions between
individuals. All individuals are initially susceptible (S), except for
one that randomly becomes infected, called patient zero. At each
time step, infected individuals (I) spread the disease to susceptible
neighbors with a probability $\beta \omega_i$, where $\beta$ is the
intrinsic infectivity of the disease, and infected individuals recover
(R) after $t_r$ time steps. We remark that bridge nodes, i.e., the
individuals present in both layers, change their state (if they become
infected or get recovered) simultaneously in both layers. The
propagation stops when the number of infected individuals is zero in
both layers. By using the model described above, we can write the
transmissibility of the disease in layer $i = A, B$ as $T_{a_i} =
\sum_{t=1}^{t_r}[(1-\beta e^{-a_i})^t -
  (1-\beta)^t]/(a_it)$~\cite{Val-13}. Note that $T_{a_i}$ is a
decreasing function of the disorder intensity $a_i$ because, since
$\omega_i \, \epsilon \, [e^{-a_i},1]$, for higher $a_i$ values
shorter contact times become more probable, and hence the disease is
less likely to propagate. In the limit $a_i \rightarrow \infty$ we
have that $T_{a_i} \rightarrow 0$, which is a complete lockdown
scenario where each individual is isolated from the rest. On the other
hand, in the limit $a_i \rightarrow 0$ there is no disorder in layer
$i$, then the infection probabilities throughout layer $i$ are all
simply equal to $\beta$, recovering the original SIR model in which
the transmissibility is $T_{a_i} \rightarrow T = 1 - (1 -
\beta)^{t_r}$. 

Setting different $a_i$ values allows us to implement a particular
social distancing intensity in each layer. Inspired by the current
events related to the COVID-19 pandemic and the policies undertaken by
authorities in many countries, we describe next the implementation of
our social distancing strategy. We assume that the population can be
divided into two layers. One layer represents people who stay in their
homes, where interactions are more intimate and usually have a
prolonged duration. The other layer represents social interactions in
environments where precaution measures are undertaken more strictly,
such as workplaces, public transport, or grocery shops. In this
structured system, it is reasonable that a fraction $q$ of individuals
may belong to both groups of people, e.g., essential workers (food
supply and distribution systems, sanitary system, public transport,
etc.). We propose that the social distancing policies undertaken in
each layer are related through their topology, setting $a_B = a_A
\kappa_B / \kappa_A$. The factor $\kappa_i = \langle k^2_i \rangle /
\langle k_i \rangle$ is the \emph{branching
  factor}~\cite{Gra-83,New-02} of layer $i = A, B$, where $\langle k_i
\rangle$ and $\langle k^2_i \rangle$ are the first and second
moments of the degree distribution $P_i(k)$, respectively. We remark
that for an ER layer the branching factor increases with the average
connectivity as $\kappa = \valmed{k} + 1$, while in a SF layer with
exponential cutoff $\kappa$ increases as the heterogeneity parameter
$\lambda$ decreases. In this way, if layer $B$ represents the social
contacts outside the households, where it is expected a more
heterogeneous distribution of the interactions, we have $\kappa_B >
\kappa_A$ and hence $a_B > a_A$; this means that a stronger distancing
policy is implemented in layer $B$ to counter the structure effects
over the spread of the disease in that layer (see
Sec.\ref{phaseR}). We remark that, in our model, the social distancing
strategy is static and is applied from the start of the propagation
of the disease, distributing fixed weights to the links. This could
represent an optimistic scenario, where the authorities of a
particular region (e.g., city or country) are well informed about the
existence of an infectious disease, and they immediately undertake
inflexible measures to get the most out of them.

Finally, we define the relevant magnitudes of the model, at the final
stage, in which we focus to study our distancing strategy. On the one
hand, $R$ is the relative size of the giant component of recovered
individuals, which is formed by all the individuals in the multiplex
network that get infected. On the other hand, GCS is the relative
size of the mutual giant component of susceptible individuals,
composed of susceptible individuals that are connected within a layer
and between the two layers of the multiplex network. Depending on the
context, we will use GCS to refer to the mutual giant component of
susceptible individuals or to its relative size. Note that the
GCS is not defined in the absence of overlap, i.e., for $q = 0$.

Considering the model presented, for the analysis of the proposed
strategy we examine disease parameters $\beta$ and $t_r$ so that the
system enters an epidemic phase without disorder, i.e., $R > 0$ for
$a_A = a_B = 0$ (thus the interactions in both layers have the largest
temporal duration). We want to analyze whether or not we can find a
pair of critical disorder intensity values $(a_{Ac},a_{Bc})$ that
prevent the disease from becoming an epidemic, and study how they
depend on the fraction $q$ of shared nodes. In addition, we look for
critical disorder intensity values $(a^*_A,a^*_B)$ that can prevent
the GCS from falling apart, which would certainly cause a significant
disruption in the economy of a region or country (a scenario
characterized by GCS$\,\, = 0$). In the next section, we develop a
theoretical approach that facilitates the analysis of the phase space
for $R$ and GCS.

\section{Theory: Results at the final stage}
\label{theory}

\subsection{Phase space for $R$}
\label{phaseR}

It has been demonstrated that in isolated networks the final stage of
the SIR model~\cite{Ker-27,Gra-83,Ander-92,New-02,Bocc-06,Pas-15} maps
exactly into link percolation~\cite{New-01,New-02,Brau-07} in which
links between nodes are occupied with probability $p$. Thus, the
relevant magnitudes of this model can be obtained theoretically. The
mapping holds in the thermodynamic limit, where $N \rightarrow
\infty$, and considering that the number of recovered individuals is
zero unless they are above a threshold $s_c$~\cite{Lag-09}, which
distinguishes between an epidemic and a small outbreak. In isolated
complex networks, the critical transmissibility for which the system
switches from an epidemic to a non-epidemic phase is $T_c = 1/(\kappa
- 1)$, where $\kappa$ is the branching factor ~\cite{Gra-83,New-02} of
the network.

\begin{figure}
  \subfloat{%
    \begin{overpic}[width=0.49\textwidth]{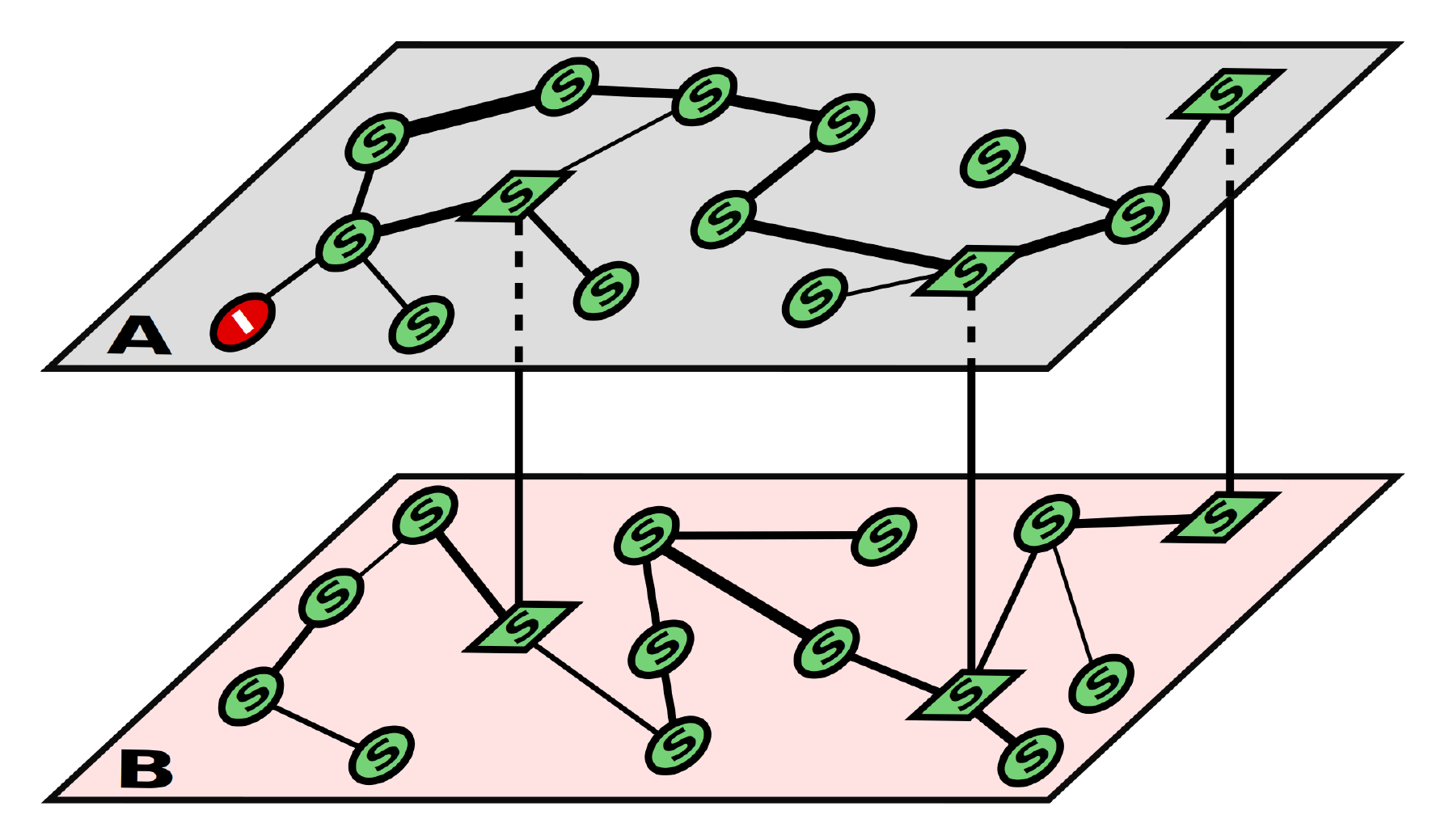}
      \put(0,51){\bf{(a)}}
    \end{overpic}
  }\hfill
  \subfloat{%
    \begin{overpic}[width=0.49\textwidth]{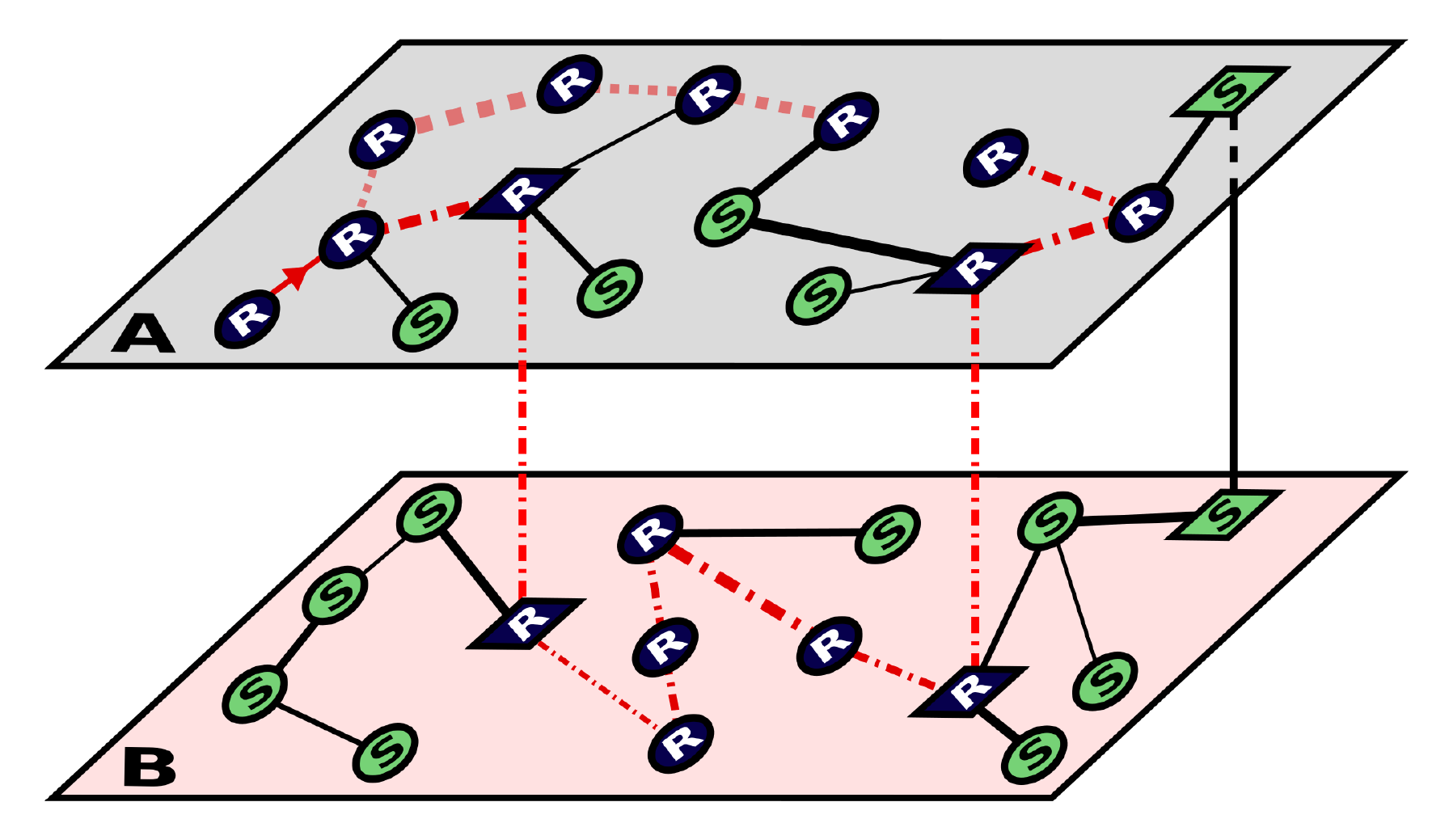}
      \put(0,51){\bf{(b)}}
    \end{overpic}
  }
  \caption{Scheme of a disordered multiplex network
    formed by two overlapped layers, $A$ and $B$. The size of
    the layers is $N_A = N_B = 15$, and the fraction of nodes present in
    both layers is $q = 3/15 = 0.2$ (vertical lines are used as a guide
    to show the shared nodes, which are represented by boxes). The
    thickness of the segments represents the diversity of the normalized
    contact times $\omega_{ij}$ between individuals. (a) Initially, all
    the individuals are in the susceptible (S) state, except for an
    infected (I) node, which kick-starts the propagation of the
    disease. (b) At the final stage, the recovered (R) individuals are
    connected by the branches of infection, which originate from the
    link denoted by a red arrow. One of the branches, denoted by dotted
    lines, corresponds to the spread of the disease only through layer
    $A$, and is represented by the first term in Eq.~(\ref{fA}). The
    other branch, denoted by dash-dotted lines, is a branch of infection
    that spreads through both layers and is represented by the second
    term in Eq.~(\ref{fA}).}\label{multiplex}
\end{figure}

Next we proceed to map the final stage of our model into
link percolation using the branching theory and the generating
functions framework~\cite{New-01,New-02,New-03,Brau-07,Brau-17}. Given
a two-layer multiplex network with overlap $0 < q \le 1$, we can write
a system of transcendental coupled equations for $f_i(T_A,T_B) \equiv
f_i$, which is the probability that a branch of infection (formed by
recovered individuals) that originates from a random link in layer $i
= A, B$ expands infinitely through any of the layers,
\begin{eqnarray}
  \label{fA}
  f_A & = (1 - q)(1 - G_1^A(1 - T_A f_A)) \, + q(1 - G_1^A(1 - T_A f_A) G_0^B(1 - T_B f_B)), \\
  \label{fB}
  f_B & = (1 - q)(1 - G_1^B(1 - T_B f_B)) \, + q(1 - G_1^B(1 - T_B f_B) G_0^A(1 - T_A f_A)),
\end{eqnarray}
where $G_0^i(x) = \sum_k P_i(k) x^k$ and $G_1^i(x) = \sum_k (k P_i(k)
  / \langle k_i \rangle) x^{k-1} = {G_0^i}'(x)/{G_0^i}'(1)$ (with $G'
\equiv dG/dx$ and $G_0'(1) = \langle k \rangle$) are the generating
functions of the degree and the excess degree distributions,
respectively~\cite{New-01,New-02,New-03,Brau-07,Brau-17}. Note that
the factor $G_1^i(x)$, with $x = 1 - T_i f_i$, is the probability that
in layer $i$ a branch of infection reaches a node with connectivity
$k$, so that it cannot keep extending through its $k - 1$ remaining
connections. In a similar way, $G_0^i(x)$ is the probability that a
randomly chosen node is not reached by a branch of infection through
its $k$ connections in layer $i$. Thus, $f_A$ is the sum of two main
terms. First, the probability of reaching an individual present only
in layer $A$ (with probability $1 - q$) so that the branch of
infection expands through any of the $k - 1$ remaining connections of
the individual in that layer, and, second, the probability of reaching
one of the shared nodes (with probability $q$) so that the branch of
infection expands through any of its $k - 1$ contacts in layer $A$ or
through any of its $k$ connections in layer $B$ (see
Fig.~\ref{multiplex}).
An analogous interpretation holds for $f_B$. Once we calculate the
non-trivial roots of Eqs.~(\ref{fA}) and ~(\ref{fB}), then the
fractions $R_A$, $R_B$, and $R$ of recovered individuals (i.e., those
reached by the branches of infection) can be obtained from
\begin{eqnarray}
  \label{recA}
  R_A & = & (1 - q)(1 - G_0^A(1 - T_A f_A)) \, + \xi_R, \\
  \label{recB}
  R_B & = & (1 - q)(1 - G_0^B(1 - T_B f_B)) \, + \xi_R, \\
  \label{rec}
  R & = & (R_A + R_B - \xi_R)/(2 - q),
\end{eqnarray}
where $\xi_R = q(1 - G_0^A(1 - T_A f_A)G_0^B(1 - T_B f_B))$ is the
fraction of shared recovered nodes at the final stage. The factor $2 -
q$ in Eq.~(\ref{rec}) accounts for the total number of individuals in
the system, which is $(2 - q)N$. In Sec.~\ref{appendix} we
show the agreement between the presented equations and the simulations
of the model. Generally, if there are critical disorder intensities
$(a_{Ac},a_{Bc})$ for a given $q$ value, these can be computed by
solving the system of equations formed by $a_{Bc} = a_{Ac} \kappa_B /
\kappa_A$ and the equation det$(J^f - I) = 0$ evaluated at $f_A = f_B
= 0$ (since at the critical point none of the branches of infection
expands infinitely). Here $I$ is the identity matrix and $J^f$ is the
Jacobian matrix of the system of Eqs.~(\ref{fA}) and~(\ref{fB}),
$J^f_{i,k} = \partial f_i / \partial f_j$, then
%
\[
J^f |_{f_A = f_B = \, 0} =
\left( {\begin{array}{cc}
     T_A (\kappa_A - 1) & q T_B \valmed{k_B} \\
     q T_A \valmed{k_A} & T_B (\kappa_B - 1) \\
  \end{array} } \right),
\]
where $\kappa_i$ and $\valmed{k_i}$ are the branching factor and the
average connectivity in layer $i = A, B$. The critical disorder
intensities $(a_{Ac},a_{Bc})$ are given by the system
\begin{eqnarray}
  \label{aBaA}
  a_{Bc} & = & a_{Ac} \frac{\kappa_B}{\kappa_A}, \\
  \label{TBc}
  T_{Bc} & = & \frac{T_{Ac} (\kappa_A - 1) - 1}{[T_{Ac} (\kappa_A - 1) - 1](\kappa_B - 1) - q^2 T_{Ac} \valmed{k_A} \valmed{k_B}},
\end{eqnarray}
which we solve numerically for $q \, \epsilon \, [0,1]$, and where
$T_{Ac} \equiv T_A(a_{Ac})$, $T_{Bc} \equiv T_B(a_{Bc})$, and $a_{Bc}
\equiv a_{Bc}(a_{Ac})$. Note that Eq.~(\ref{TBc}) differs from the one
obtained in Ref.~\cite{Buo-14}, where both layers of the multiplex
network have the same transmissibility $T$, which yields a quadratic
equation for $T_c$ (the critical transmissibility) with only one
stable solution. The result expressed by Eq.~(\ref{TBc}) reflects the
case where no relation between the disorder intensities is
implemented, meaning that for every $a_A$ value could be a critical
value $a_{Bc}$, maintaining a fixed overlap $q$.

\begin{figure}
  \subfloat{%
    \begin{overpic}[width=0.49\textwidth]{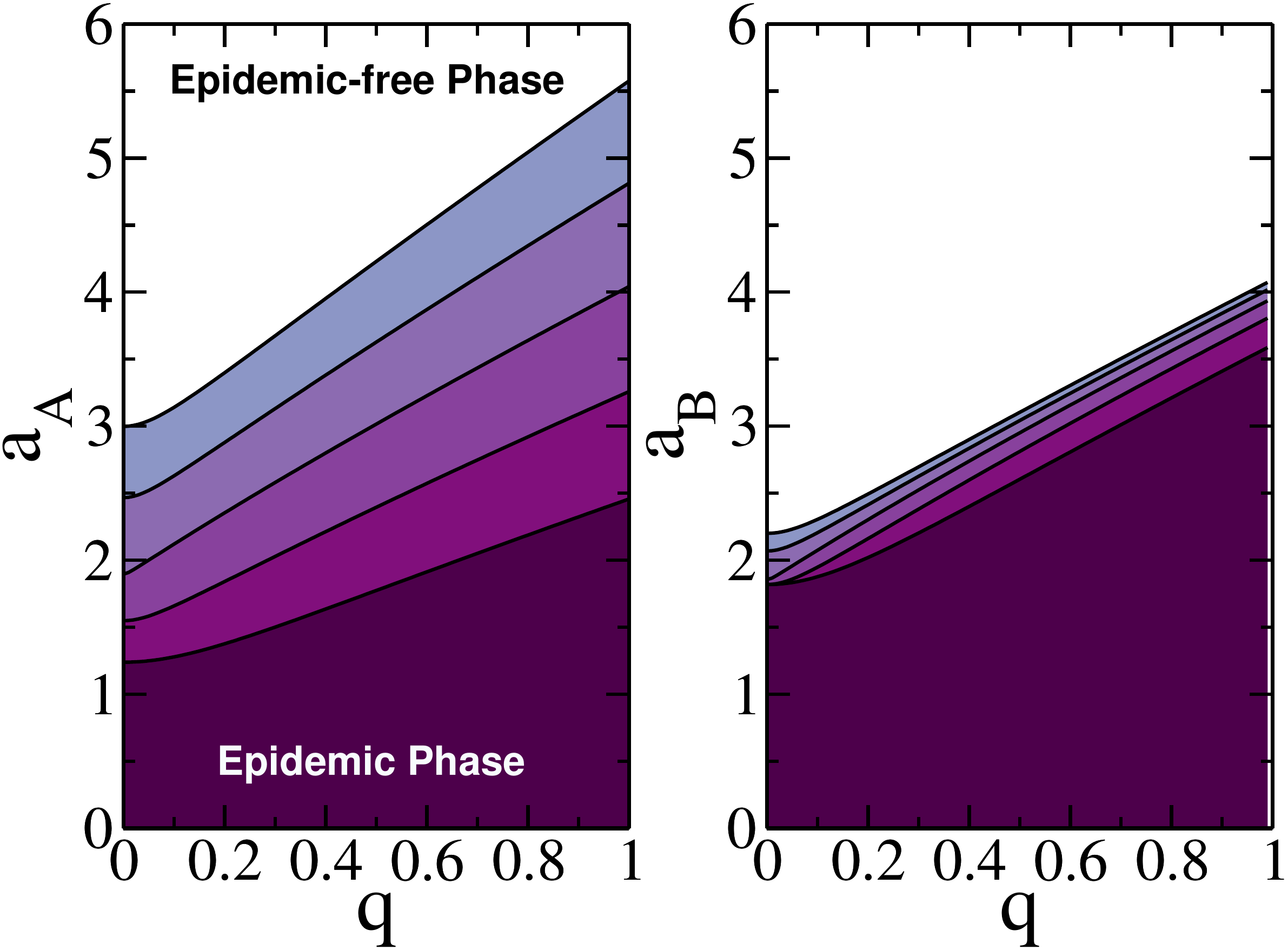}
      \put(50.5,0){\bf{(a)}}
    \end{overpic}
  }\hfill
  \subfloat{%
    \begin{overpic}[width=0.49\textwidth]{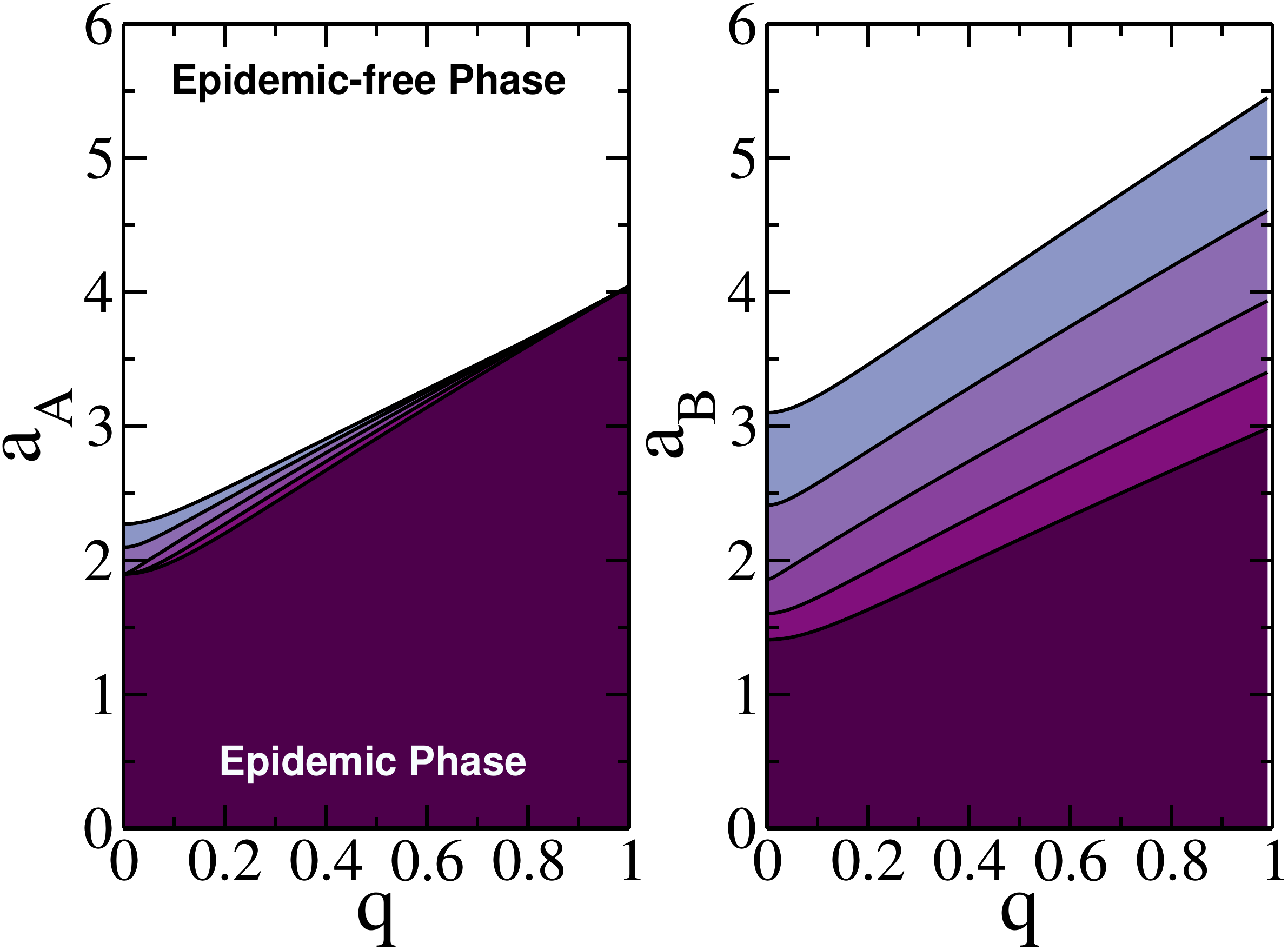}
      \put(50.5,0){\bf{(b)}}
    \end{overpic}
  }
  \caption{Phase diagrams for $R$ on the
    $(q,a_A)$ and $(q,a_B)$ planes. (a) Diagrams for an ER layer with
    different values of $\valmed{k_A}$ and a SF layer with $\lambda_B
    = 2.5$ and cutoff $k^B_c = 20$. The solid and black curves
    represent the critical disorder intensity $a_{Ac}$ (left-hand) and
    $a_{Bc}$ (right-hand) for $\valmed{k_A} = 3, 4, 5, 6, 7$, from
    bottom to top. (b) Diagrams for an ER layer wih $\valmed{k_A} = 5$
    and a SF layer with cutoff $k^B_c = 20$ and $\lambda_B = 2.1, 2.3,
    2.5, 2.7, 2.9$, from top to bottom. Colored regions below the
    critical curves indicate epidemic phases ($R > 0$), which expand
    as the overlap $q$ increases. Note that in (a) the increase of
    $\valmed{k_A}$ causes a major increment on the critical
    intensities of layer $A$, while in $B$ the increment is smaller. A
    similar effect occurs in (b), when decreasing $\lambda_B$. The
    remaining parameters of the layers are $k^A_{\rm min} = 0$,
    $k^A_{\rm max} = 20$, $k^B_{\rm min} = 2$, and $k^B_{\rm max} = 250$.}\label{phaseR-ER-SFT}
\end{figure}

In Fig.~\ref{phaseR-ER-SFT} we show phase diagrams for $R$ on the
$(q,a_A)$ and $(q,a_B)$ planes, where we use two layers with different
degree distributions. The selected parameters for the disease are
$\beta = 0.1$ and $t_r = 5$.
In Fig.~\ref{phaseR-ER-SFT} (a) we use an ER layer with variable
$\valmed{k_A}$ and a SF layer with $\lambda_B = 2$ and cutoff $k^B_c =
20$. On the left side of the figure, the curves show the critical
values $a_{Ac}$ as a function of the overlap $q$ between the layers,
for the different values of $\valmed{k_A}$. The colored regions
correspond to the epidemic phase, where $R > 0$, for a given
$\valmed{k_A}$ value, while above the critical curves the system
presents an epidemic-free phase, where $R = 0$. The corresponding
critical values $a_{Bc}$ for the disorder intensity in layer $B$ are
plotted alongside, considering
Eq.~(\ref{aBaA}). Fig.~\ref{phaseR-ER-SFT} (b) shows the critical
curves for an ER layer with $\valmed{k_A} = 4$ and a SF layer with
variable $\lambda_B$. As expected in all the cases, the increase of
the overlap $q$ causes the epidemic regions to expand, since the
spread of the disease is favored by the individuals that propagate it
through both layers. This means that the distancing needed to halt the
epidemic must be stronger in both layers, thus decreasing the average
duration of the interactions. We also observe that the increase of
$\kappa_i$ (due to the increase of $\valmed{k_A}$ in
Fig.~\ref{phaseR-ER-SFT} (a) or the decrease of $\lambda_B$ in
Fig.~\ref{phaseR-ER-SFT} (b) causes a major increment of the critical
intensities in layer $i$, while in the other layer the effect is
smaller. In these cases, if the structure of a layer changes in favor
of the dissemination of the disease (i.e., increase in $\kappa_i$)
while the other remains unchanged, much of the additional distancing
necessary to avoid an epidemic falls on the altered
layer.



\subsection{Phase space for GCS}

In what follows we present a set of equations that allow us to
calculate the size GCS of the mutual giant component of susceptible 
individuals of the multiplex network at the final stage of the
process. It is straightforward to write GCS$_A$ and GCS$_B$, the
fraction of nodes that belong to the giant components of susceptible
individuals in layer $A$ and $B$ (i.e., the susceptible individuals
connected within each one of the layers), respectively,
\begin{eqnarray}
  \label{GCSA}
  GCS_A & = (1 - q)(G_0^A(1 - T_A f_A) - G_0^A(\nu_A)) \, + \xi_S,
  \\
  \label{GCSB}
  GCS_B & = (1 - q)(G_0^B(1 - T_B f_B) - G_0^B(\nu_B)) \, + \xi_S.
\end{eqnarray}
The first terms takes into account the probability that a node that is
part of only one of the layers (with probability $1 - q$) belongs to
the giant component of that layer. This can be written as the
probability that a node of layer $i$ is susceptible ($G_0^i(1 - T_i
f_i)$) minus the probability of the node being susceptible but not
belonging to the giant component ($G_0^i(\nu_i)$). On the other hand,
$\xi_S = q(G_0^A(1 - T_A f_A) G_0^B(1 - T_B f_B) - G_0^A(\nu_A) G_0^B
(\nu_B))$ is the fraction of shared nodes that belong to the GCS (the
mutual giant component). A node is susceptible and does not belong to
the GCS if none of its links lead to susceptible nodes that do belong
to the GCS. But, if one of these links connects to an R node, in order
to be susceptible, the node cannot have been infected by this R
node. Thus, we define $\nu_i$ as the probability that a random link
from layer $i$ leads to a susceptible node that does not belong to the
GCS, or to an R node. However, note that in the last case the link
must be unoccupied, with probability $1 - T_i$. Note that similarly to
Eqs.~(\ref{fA}) and~(\ref{fB}), we can write a set of coupled
transcendental equations for the probabilities $\nu_A (T_A,T_B) \equiv
\nu_A$ and $\nu_B (T_A,T_B) \equiv \nu_B$,
\begin{eqnarray}
  \label{nuA}
  \nu_A & = (1 - T_A)f_A + (1 - q)G_1^A (\nu_A) + qG_1^A (\nu_A) G_0^B
  (\nu_B), \\
  \label{nuB}
  \nu_B & = (1 - T_B)f_B + (1 - q)G_1^B (\nu_B) + qG_1^B (\nu_B) G_0^A
  (\nu_A).
\end{eqnarray}
From left to right, the first term is the probability that a random
link in layer $i$ leads to a recovered node, which is $f_i$ but
considering that the link is unoccupied, with probability $1 -
T_i$. The second is the probability that the random link connects to a
node that belongs only to layer $i$ (with probability $1 - q$), so
that none of its outgoing links lead to susceptible nodes belonging to
the GCS, nor any of its outgoing links leads to a recovered node and
is occupied. Finally, the last term is similar to the second, but the
random link in layer $i$ leads to a shared node (with probability $q$)
so that besides its outgoings links in layer $i$, none of its links in
layer $j$ connect to susceptible nodes that are part of the GCS, and
none of its links connects to a recovered node and is occupied. Once
the values of $\nu_i$ and consequently GCS$_i$ for $i = A, B$ are
obtained from Eqs.~(\ref{GCSA}-\ref{nuB}), the size of the GCS can be
computed as
\begin{equation}
  \label{GCS}
  GCS = (GCS_A + GCS_B - \xi_S)/(2 - q).
\end{equation}
In Sec.~\ref{appendix} we show the agreement between the equations
presented and the simulations of the model.

To study the phase space for the GCS, we define $\mu_i$ as the
probability that a random link in layer $i = A, B$ connects to a node
belonging to the GCS (similarly to what was done in Sec.~\ref{phaseR}
with $f_i$ for the recovered individuals). Recall that $\nu_i$ is the
probability that a random link connects to an S node which does not
belong to the GCS or that it connects to an R node but considering
that the link is unoccupied with probability $1 - T_i$. Then we have
that $\mu_i = 1 - (\nu_i + T_i f_i)$, and we obtain a system of
equations for $\mu_A$ and $\mu_B$:
\begin{eqnarray}
  \label{muA}
  \mu_A & = 1 - f_A - (1 - q)G_1^A(u_A - \mu_A) - qG_1^A(u_A - \mu_A)
  G_0^B(u_B - \mu_B), \\
  \label{muB}
  \mu_B & = 1 - f_B - (1 - q)G_1^B(u_B - \mu_B) - qG_1^B(u_B - \mu_B)
  G_0^A(u_A - \mu_A), 
\end{eqnarray}
where $u_i \equiv 1 - T_i f_i$. If there are critical disorder
intensities $(a^*_A,a^*_B)$ for a given $q$ value, these can be
computed by solving the system of equations formed by $a^*_B = a^*_A
\kappa_B / \kappa_A$, the equation det$(J^{\mu} - I) = 0$ evaluated
at $\mu_A = \mu_B = 0$ and Eqs.~(\ref{fA}) and~(\ref{fB}) for $f_A$
and $f_B$, respectively. Here $J^{\mu}$ is the Jacobian matrix of the
system of Eqs.~(\ref{muA}) and~(\ref{muB}), $J^{\mu}_{i,k} = \partial
\mu_i / \partial \mu_j$, then 
\[
J^{\mu} |_{\mu_A = \mu_B = \, 0} =
\left( {\begin{array}{cc}
     (1 - q) {G_1^A}'(u_A) + q {G_1^A}'(u_A) G_0^B(u_B) & q G_1^A(u_A)
     G_1^B(u_B) \valmed{k_B} \\
     q G_1^B(u_B) G_1^A(u_A) \valmed{k_A} & (1 - q) {G_1^B}'(u_B) + q
     {G_1^B}'(u_B) G_0^A(u_A) \\
  \end{array} } \right),
\]
where $u_A \equiv u_A(a^*_A,f_A)$, $u_B \equiv u_B(a^*_B,f_B)$, and
$a^*_B \equiv a^*_B(a^*_A)$. We solve this system numerically for $q
\, \epsilon \, (0,1]$, since the GCS (the mutual giant component of
  susceptible individuals) is not defined in the absence of overlap,
  i.e., for $q = 0$.
  
\begin{figure}
  \subfloat{%
    \begin{overpic}[width=0.49\textwidth]{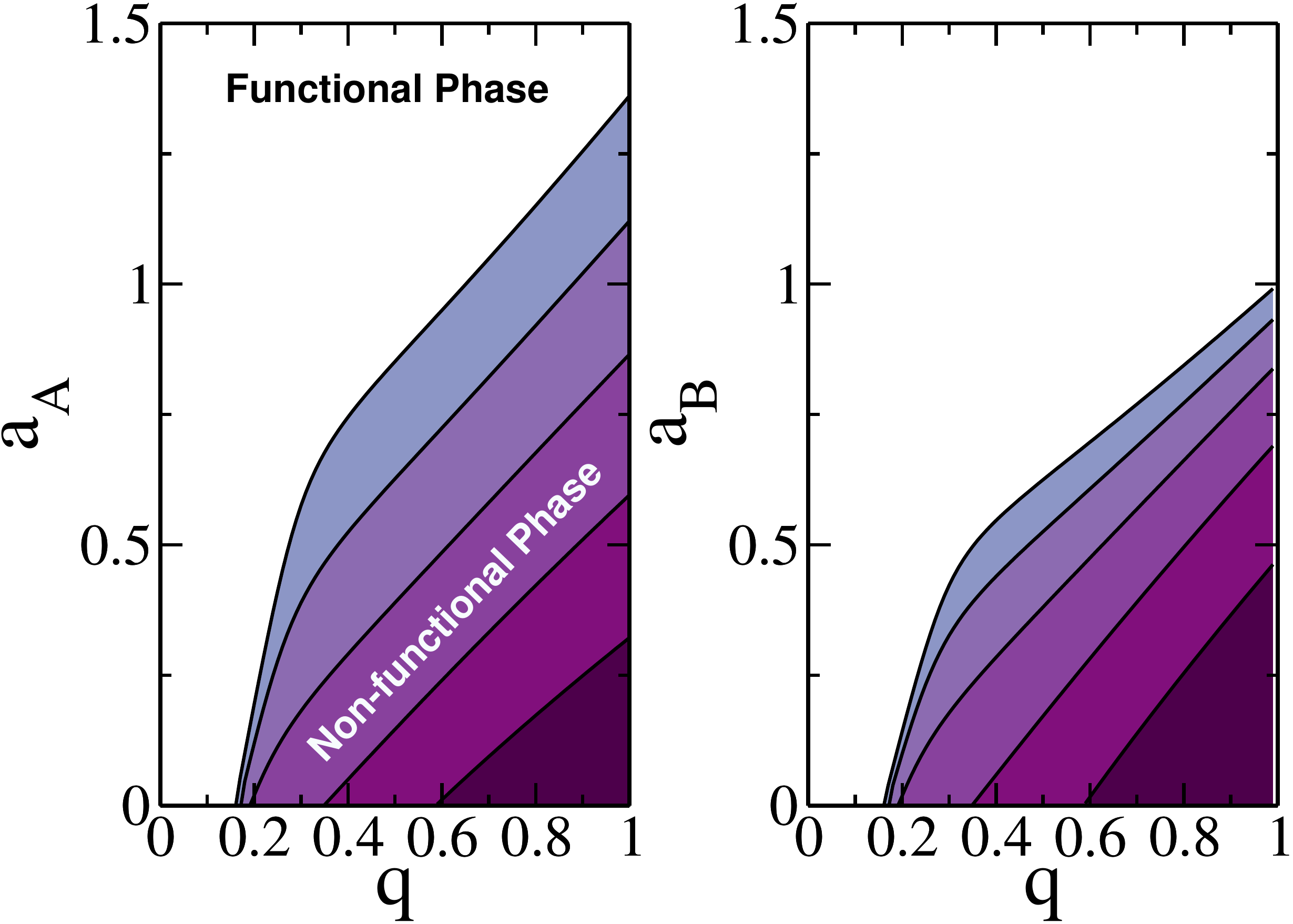}
      \put(51.5,0){\bf{(a)}}
    \end{overpic}
  }\hfill
  \subfloat{%
    \begin{overpic}[width=0.49\textwidth]{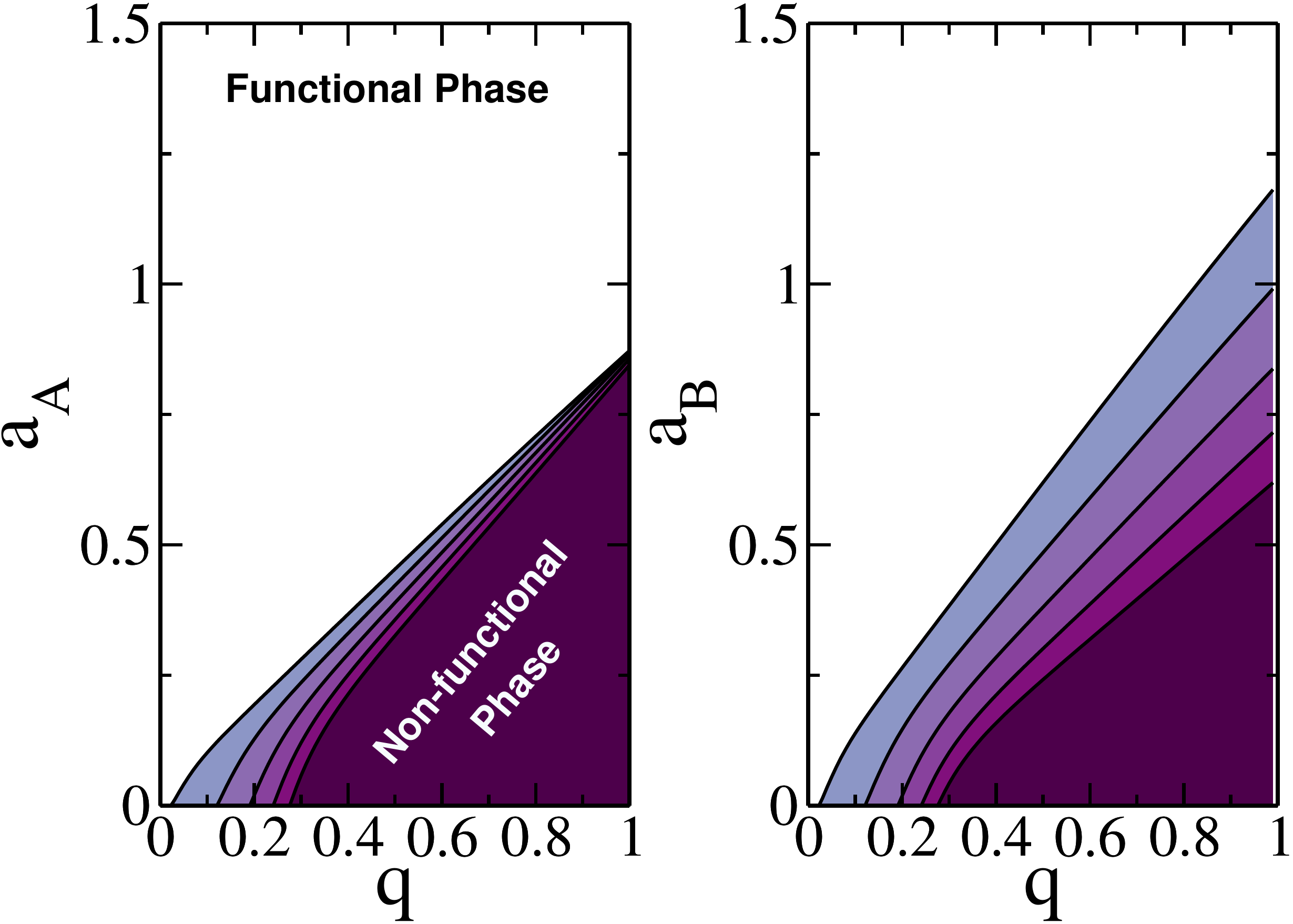}
      \put(51.5,0){\bf{(b)}}
    \end{overpic}
  }
  \caption{Phase diagrams for GCS on the
    $(q,a_A)$ and $(q,a_B)$ planes. (a) Diagrams for an ER layer with
    different values of $\valmed{k_A}$ and a SF layer with $\lambda_B
    = 2.5$ and cutoff $k^B_c = 20$. The solid and black curves
    represent the critical disorder intensity $a^*_A$ (left-hand) and
    $a^*_B$ (right-hand) for $\valmed{k_A} = 3, 4, 5, 6, 7$, from
    bottom to top. (b) Diagrams for an ER layer wih $\valmed{k_A} = 5$
    and a SF layer with cutoff $k^B_c = 20$ and $\lambda_B = 2.1, 2.3,
    2.5, 2.7, 2.9$, from top to bottom. Colored regions below the
    critical curves indicate non-functional phases (GCS$\,\, = 0$), and
    functional phases (GCS$\,\, > 0$) above the curves. Unlike the results
    shown in the phase diagrams for $R$, in this case the critical
    intensities may vanish, i.e., $a^*_A = a^*_B = 0$, if the overlap
    $q$ is relatively small. This means that the functionality of the
    GCS is ensured, independently of the intensity of the implemented
    distancing strategy.}\label{phaseGCS-ER-SFT}
\end{figure}

In Fig.~\ref{phaseGCS-ER-SFT} we show phase diagrams for GCS on the
$(q,a_A)$ and $(q,a_B)$ planes, where we use two layers with different
degree distributions (the same layers that we used for $R$ in
Fig.~\ref{phaseR-ER-SFT}). The selected parameters for the disease are
$\beta = 0.1$ and $t_r = 5$.
In Fig.~\ref{phaseGCS-ER-SFT} (a) we use an ER layer with variable
$\valmed{k_A}$ and a SF layer with $\lambda_B = 2$ and cutoff $k^B_c =
20$. On the left side of the figure, the curves show the critical
values $a^*_A$ as a function of the overlap $q$ between the layers,
for the different values of $\valmed{k_A}$. The colored regions
correspond to the non-functional phase, where GCS$\,\, = 0$, for a given
$\valmed{k_A}$ value, while above the critical curves the system
presents a functional phase, where GCS$\,\, > 0$. The corresponding
critical values $a^*_B$ for the disorder intensity in layer $B$ are
plotted alongside, considering
Eq.~(\ref{aBaA}). Fig.~\ref{phaseGCS-ER-SFT} (b) shows the critical
curves for an ER layer with $\valmed{k_A} = 4$ and a SF layer with
variable $\lambda_B$. As we observe, there are similarities between
these results and the phase diagrams presented for $R$. On the one
hand, the increase of $\kappa_i$ causes a major increment in layer $i$
of the distancing necessary to protect the GCS, while in the remaining
layer the change is smaller. However, we note that there is a range
of $q$ values in which the critical disorder intensities vanish, i.e.,
$a^*_A = a^*_B = 0$. This range depends on the parameters of each
particular multiplex network and decreases with $\kappa_i$, as can be
seen in Figs.~\ref{phaseGCS-ER-SFT} (a) and (b). In this region, the
GCS exists even when a distancing strategy is not implemented in any
layer, thus the system does not lose its functionality. For values of
the overlap $q$ that exceed this range, the system needs a minimum
distancing to avoid the collapse of the GCS, and an increase of $q$
causes the non-functional regions to expand.

\begin{figure} 
  \centering
  \includegraphics[width=0.49\textwidth]{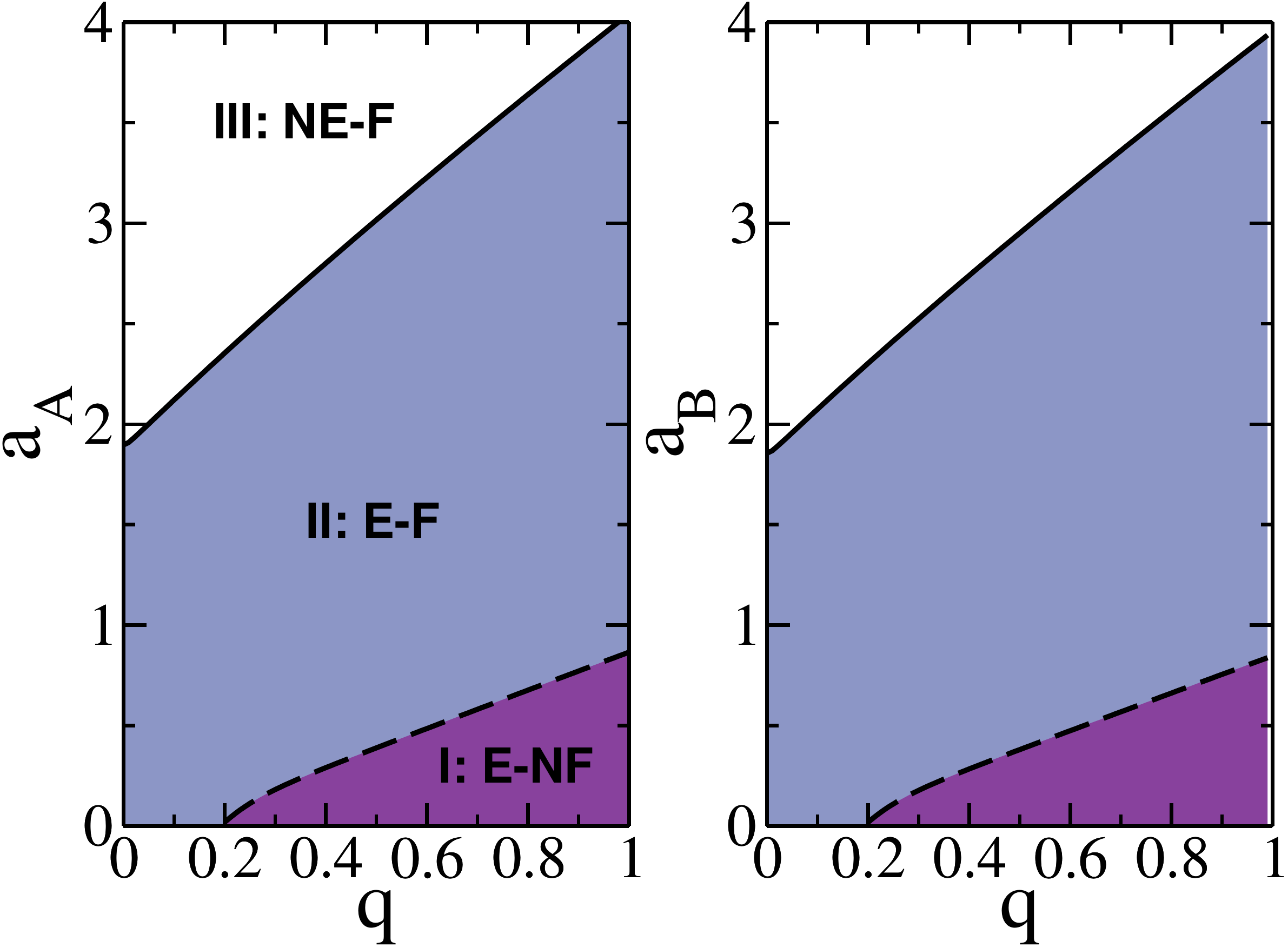}
  \caption{\label{strategy} Typical outcomes of the social
    distancing strategy applied to a multiplex network. Solid and
    dashed lines represent the critical curves for $R$ and GCS,
    respectively. In region I of the diagram the social distancing
    policies are rather weak, leading not only to an epidemic, but
    also to the crumbling of the GCS, which disrupts the functionality
    of the system (E-NF phase). As the policies are intensified, in
    region II, the GCS is prevented from falling apart, while the
    emergence of an epidemic becomes less likely (E-F phase). In
    region III, a strict enough social distancing policy (i.e., high
    disorder intensities in both layers) ensures the system lies on
    a non-epidemic and functional phase (NE-F phase), which is the
    best possible scenario. For the presented results, we use an ER
    layer with $\valmed{k_A} = 5$ and a SF layer with $\lambda_B =
    2.5$ and cutoff $k^B_c = 20$. Also, we consider a disease with
    $\beta = 0.1$ and $t_r = 5$.}
\end{figure}

Next we present the summarized results of the social distancing
strategy proposed. Looking at an individual case of the multiplex
networks previously presented, in Fig.~\ref{strategy} we show the
critical curves for $R$ and GCS, the sizes of the giant components
of susceptible and recovered individuals throughout the entire system,
respectively.
We observe now that the phase diagram is divided into three
regions. In region I the disease is certainly likely to spread because
the disorder intensities implemented in both layers are quite low,
which not only fails to prevent an epidemic but also makes the GCS
fall apart, causing the system to collapse (E-NF phase). It may happen
that even in the absence of a distancing strategy the system does not
lose the functionality (i.e., GCS$\,\, > 0$ for $a_A = a_B = 0$). This
depends on the structure of the multiplex network system (the
$\kappa_B/\kappa_A$ relation) and occurs only for relatively small
values of the overlap $q$ (in Fig.~\ref{strategy}, for $q < 0.2$
approximately). As distancing increases, in region II the epidemic
still can not be avoided, but there emerges a GCS, which means that
the system remains functional (E-F phase). Finally, region I
corresponds to a non-epidemic and functional phase (NE-F), the best
possible scenario, in which the disorder intensities in both layers are
high enough, i.e., social distancing measures are undertaken
intensively. In this region, the system avoids an epidemic and
maintains its functionality. Considering this, even though the social
distancing efforts may not prevent that a particular disease extends
through a significant portion of the population (regions I and II in
Fig.~\ref{strategy}), they could serve to the purpose of keeping the
integrity of the system (regions II and III). It is important to note
that, if a GCS remains functional after the end of an epidemic, it is
highly recommended not to relax the set of measures undertaken to
prevent the transmission of the disease, since there is always the
possibility of a second outbreak (originating, for instance, from an
imported or undetected case).

\bigskip

Looking forward, we comment on a few elements that we are interested
in analyzing in future research. For instance, it is known that
nowadays passenger traffic is one of the main causes of the
dissemination of a disease across different regions (cities, states,
and countries). One way to tackle this issue is to isolate individuals
once they arrive to its destination, which is currently being
implemented during the COVID-19 pandemic. In this way, it is expected
that the passengers cannot spread the disease within the region. To
take into account this possibility, instead of using an overlapped
multiplex network, we could devise a model in which shared nodes do
not belong to both layers, but rather they connect to nodes from other
layers according to an inter-layer degree distribution. These nodes
would represent individuals that travel to other regions or countries,
which are isolated with a certain probability. These considerations
may help in deepening the understanding of these processes,
encouraging researchers to devise more suitable and efficient
strategies for preventing and mitigating the spread of a disease. In
addition, it would be relevant to study the temporal evolution of a
disease in this scenario, and how it would respond to mitigation
measures that are undertaken with some delay.

\section{Correspondence between theoretical results and simulations}
\label{appendix}

\begin{figure}
  \subfloat{%
    \begin{overpic}[width=0.45\textwidth]{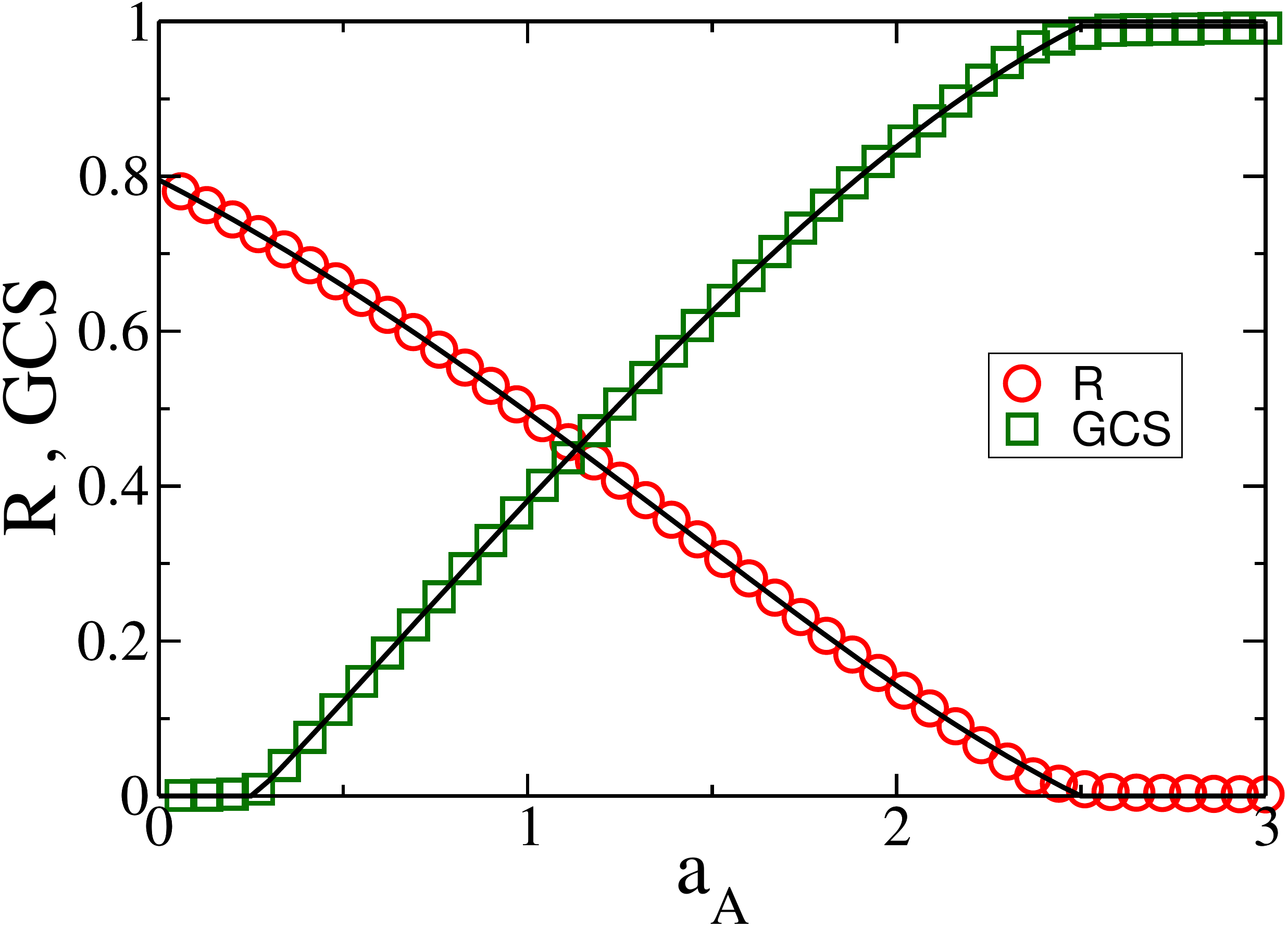}
      \put(14,64){\bf{(a)}}
    \end{overpic}
  }\hfill
  \subfloat{%
    \begin{overpic}[width=0.45\textwidth]{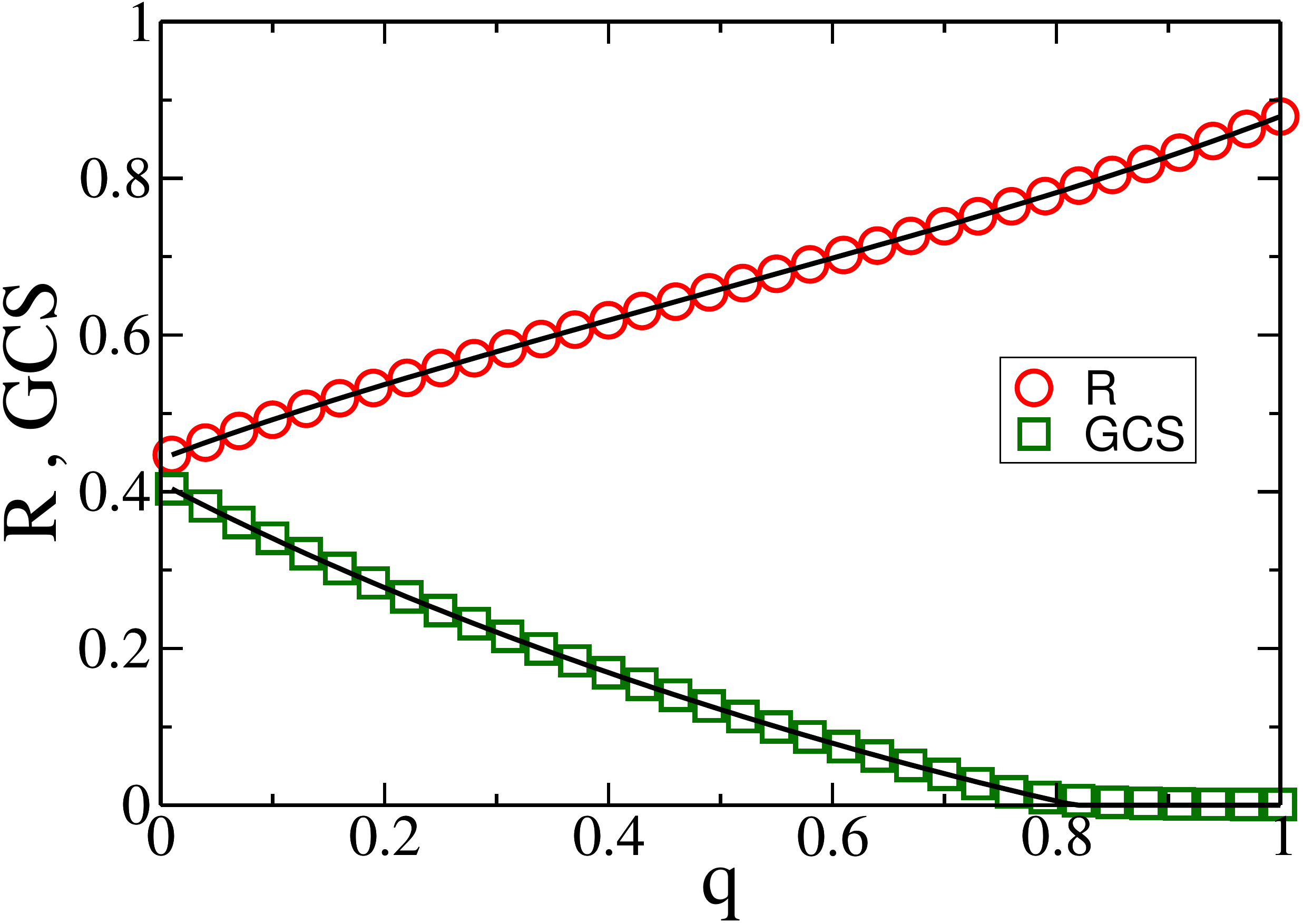}
      \put(14,63){\bf{(b)}}
    \end{overpic}
  }
  \caption{Agreement between theoretical and
    simulated results. Solid and black lines correspond to theoretical
    results, while simulations are shown in symbols. (a) Fraction $R$
    of recovered individuals (red circles) and the size GCS of the
    mutual giant component of susceptible individuals (green squares)
    as functions of the disorder intensity $a_A$ in layer $A$, for $q
    = 0.5$. The critical disorder intensities are $a_{Ac} \approx
    2.5$, $a_{Bc} \approx 4.5$ for $R$ and $a^*_A \approx 0.26 $,
    $a^*_B \approx 0.46$ for GCS. (b) $R$ and GCS as functions of
    the overlap $q$ between the layers of the multiplex network, for
    $a_A = 0.5$. We observe that for $q \approx 0.81$ the GCS
    collapses, which accounts for the overlap effect over the spread
    of the disease. For the simulations, we used two layers of size $N
    = 10^5$. $P_A(k)$ is a Poisson distribution with $\langle k_A
    \rangle = 4$, $k^A_{\rm min} = 0$ and $k^A_{\rm max} = 40$, while
    $P_B(k)$ is a power-law distribution with cutoff $k_C = 20$,
    $\lambda = 2$, $k^B_{\rm min} = 2$ and $k^B_{\rm max} = 250$.}\label{sim&theo}
\end{figure}

In this brief section we present a selected set of results from the
stochastic simulations of the model established in Sec.~\ref{model},
at the final stage, along with the results computed from the equations
obtained in Sec.~\ref{theory}. Fig.~\ref{sim&theo} shows the total
fraction $R$ of recovered individuals and the size GCS of the mutual
giant component of susceptible individuals that span the entire
multiplex network as functions of the disorder intensity $a_A$ in
layer $A$ (Fig.~\ref{sim&theo} (a)) and the overlap $q$
(Fig.~\ref{sim&theo} (b)). 
The solid and black lines correspond to the theoretical results, which
agree with the simulations that are represented by symbols. We run
$10^4$ realizations of the simulations and averaged the results
considering only realizations where the number of recovered
individuals in each layer was above a threshold $s_c =
200$~\cite{Lag-09}.

\section{Conclusions}

We apply the SIR model to study the spread of a disease in an
overlapped multiplex network composed of two layers with its own
distribution of contact times, in which a fraction $q$ of individuals
belong to both layers. We propose a social distancing strategy that
reduces the average contact time of the interactions between
individuals within each layer. This is achieved by increasing the
respective disorder intensities, which are related through the topology
of the layers. We find that as the intensity of the distancing
increases, the system can go through different phases. For low levels
of distancing, the system fails to prevent an epidemic and to protect
the functional network. The increment of the measures reduces the
spread of the disease, thus taking the system to a functional phase
(which is fundamental to keep running the economy of a society) but in
an epidemic scenario. Finally, the system reaches an epidemic-free and
functional phase for high enough levels of distancing. The critical
values increase with the overlap $q$ because the individuals that are
shared by the layers ease the spread of the disease, so that more
social distancing measures must be undertaken to prevent an
epidemic. However, depending on the structure of the multiplex network
system, relatively small values of $q$ might allow the system to
maintain the functionality, even in the absence of distancing
policies. All in all, the control of contact times between individuals
can serve as a prevention strategy that overcomes the overlap effect
in multiplex networks, preventing not only an epidemic, but also the
economic collapse of a region or country, which might be equally
harmful for society.

\acknowledgments

We acknowledge UNMdP and CONICET (PIP 00443/2014) for financial
support. C.E.L., M.A.D. and I.A.P. acknowledge CONICET for financial
support. Work at Boston University is supported by NSF Grant
No. PHY-1505000 and by DTRA Grants No. HDTRA-1-19-1-0017 and
No. HDTRA1-19-1-0016.


\begin{thebibliography}{56}%
\makeatletter
\providecommand \@ifxundefined [1]{%
 \@ifx{#1\undefined}
}%
\providecommand \@ifnum [1]{%
 \ifnum #1\expandafter \@firstoftwo
 \else \expandafter \@secondoftwo
 \fi
}%
\providecommand \@ifx [1]{%
 \ifx #1\expandafter \@firstoftwo
 \else \expandafter \@secondoftwo
 \fi
}%
\providecommand \natexlab [1]{#1}%
\providecommand \enquote  [1]{``#1''}%
\providecommand \bibnamefont  [1]{#1}%
\providecommand \bibfnamefont [1]{#1}%
\providecommand \citenamefont [1]{#1}%
\providecommand \href@noop [0]{\@secondoftwo}%
\providecommand \href [0]{\begingroup \@sanitize@url \@href}%
\providecommand \@href[1]{\@@startlink{#1}\@@href}%
\providecommand \@@href[1]{\endgroup#1\@@endlink}%
\providecommand \@sanitize@url [0]{\catcode `\\12\catcode `\$12\catcode
  `\&12\catcode `\#12\catcode `\^12\catcode `\_12\catcode `\%12\relax}%
\providecommand \@@startlink[1]{}%
\providecommand \@@endlink[0]{}%
\providecommand \url  [0]{\begingroup\@sanitize@url \@url }%
\providecommand \@url [1]{\endgroup\@href {#1}{\urlprefix }}%
\providecommand \urlprefix  [0]{URL }%
\providecommand \Eprint [0]{\href }%
\providecommand \doibase [0]{https://doi.org/}%
\providecommand \selectlanguage [0]{\@gobble}%
\providecommand \bibinfo  [0]{\@secondoftwo}%
\providecommand \bibfield  [0]{\@secondoftwo}%
\providecommand \translation [1]{[#1]}%
\providecommand \BibitemOpen [0]{}%
\providecommand \bibitemStop [0]{}%
\providecommand \bibitemNoStop [0]{.\EOS\space}%
\providecommand \EOS [0]{\spacefactor3000\relax}%
\providecommand \BibitemShut  [1]{\csname bibitem#1\endcsname}%
\let\auto@bib@innerbib\@empty
\bibitem [{\citenamefont {{World Health
  Organization}}(2020{\natexlab{a}})}]{WHO-20}%
  \BibitemOpen
  \bibfield  {author} {\bibinfo {author} {\bibnamefont {{World Health
  Organization}}},\ }\href
  {https://www.who.int/dg/speeches/detail/who-director-general-s-opening-remarks-at-the-media-briefing-on-covid-19---11-march-2020}
  {\bibinfo {title} {{WHO} {D}irector-{G}eneral's opening remarks at the media
  briefing on {COVID}-19, 11 {M}arch 2020}} (\bibinfo {year}
  {2020}{\natexlab{a}})\BibitemShut {NoStop}%
\bibitem [{\citenamefont {Cucinotta}\ and\ \citenamefont
  {Vanelli}(2020)}]{Cuc-20}%
  \BibitemOpen
  \bibfield  {author} {\bibinfo {author} {\bibfnamefont {D.}~\bibnamefont
  {Cucinotta}}\ and\ \bibinfo {author} {\bibfnamefont {M.}~\bibnamefont
  {Vanelli}},\ }\bibfield  {title} {\bibinfo {title} {{WHO} {D}eclares
  {COVID}-19 a {P}andemic},\ }\href {https://doi.org/10.23750/abm.v91i1.9397}
  {\bibfield  {journal} {\bibinfo  {journal} {Acta Bio Medica Atenei
  Parmensis}\ }\textbf {\bibinfo {volume} {91}},\ \bibinfo {pages} {157}
  (\bibinfo {year} {2020})}\BibitemShut {NoStop}%
\bibitem [{\citenamefont {Bailey}(1975)}]{Bai-75}%
  \BibitemOpen
  \bibfield  {author} {\bibinfo {author} {\bibfnamefont {N.~T.~J.}\
  \bibnamefont {Bailey}},\ }\href@noop {} {\emph {\bibinfo {title} {{The
  Mathematical Theory of Infectious Diseases}}}}\ (\bibinfo  {publisher}
  {Griffin, London},\ \bibinfo {year} {1975})\BibitemShut {NoStop}%
\bibitem [{\citenamefont {Anderson}\ and\ \citenamefont
  {May}(1992)}]{Ander-92}%
  \BibitemOpen
  \bibfield  {author} {\bibinfo {author} {\bibfnamefont {R.~M.}\ \bibnamefont
  {Anderson}}\ and\ \bibinfo {author} {\bibfnamefont {R.~M.}\ \bibnamefont
  {May}},\ }\href@noop {} {\emph {\bibinfo {title} {{Infectious Diseases of
  Humans: Dynamics and Control}}}}\ (\bibinfo  {publisher} {Oxford University
  Press, Oxford},\ \bibinfo {year} {1992})\BibitemShut {NoStop}%
\bibitem [{\citenamefont {Gonzalez}\ \emph {et~al.}(2008)\citenamefont
  {Gonzalez}, \citenamefont {Hidalgo},\ and\ \citenamefont
  {Barabasi}}]{Gon-08}%
  \BibitemOpen
  \bibfield  {author} {\bibinfo {author} {\bibfnamefont {M.~C.}\ \bibnamefont
  {Gonzalez}}, \bibinfo {author} {\bibfnamefont {C.~A.}\ \bibnamefont
  {Hidalgo}},\ and\ \bibinfo {author} {\bibfnamefont {A.-L.}\ \bibnamefont
  {Barabasi}},\ }\bibfield  {title} {\bibinfo {title} {Understanding individual
  human mobility patterns},\ }\href {https://doi.org/10.1038/nature06958}
  {\bibfield  {journal} {\bibinfo  {journal} {Nature (London)}\ }\textbf
  {\bibinfo {volume} {453}},\ \bibinfo {pages} {779} (\bibinfo {year}
  {2008})}\BibitemShut {NoStop}%
\bibitem [{\citenamefont {Cattuto}\ \emph {et~al.}(2010)\citenamefont
  {Cattuto}, \citenamefont {den Broeck}, \citenamefont {Barrat}, \citenamefont
  {Colizza}, \citenamefont {Pinton},\ and\ \citenamefont
  {Vespignani}}]{Catt-10}%
  \BibitemOpen
  \bibfield  {author} {\bibinfo {author} {\bibfnamefont {C.}~\bibnamefont
  {Cattuto}}, \bibinfo {author} {\bibfnamefont {W.~V.}\ \bibnamefont {den
  Broeck}}, \bibinfo {author} {\bibfnamefont {A.}~\bibnamefont {Barrat}},
  \bibinfo {author} {\bibfnamefont {V.}~\bibnamefont {Colizza}}, \bibinfo
  {author} {\bibfnamefont {J.-F.}\ \bibnamefont {Pinton}},\ and\ \bibinfo
  {author} {\bibfnamefont {A.}~\bibnamefont {Vespignani}},\ }\bibfield  {title}
  {\bibinfo {title} {Dynamics of person-to-person interactions from distributed
  rfid sensor networks},\ }\href {https://doi.org/10.1371/journal.pone.0011596}
  {\bibfield  {journal} {\bibinfo  {journal} {PLoS ONE}\ }\textbf {\bibinfo
  {volume} {5}},\ \bibinfo {pages} {e11596} (\bibinfo {year}
  {2010})}\BibitemShut {NoStop}%
\bibitem [{\citenamefont {Boccaletti}\ \emph {et~al.}(2006)\citenamefont
  {Boccaletti}, \citenamefont {Latora}, \citenamefont {Moreno}, \citenamefont
  {Chavez},\ and\ \citenamefont {Hwang}}]{Bocc-06}%
  \BibitemOpen
  \bibfield  {author} {\bibinfo {author} {\bibfnamefont {S.}~\bibnamefont
  {Boccaletti}}, \bibinfo {author} {\bibfnamefont {V.}~\bibnamefont {Latora}},
  \bibinfo {author} {\bibfnamefont {Y.}~\bibnamefont {Moreno}}, \bibinfo
  {author} {\bibfnamefont {M.}~\bibnamefont {Chavez}},\ and\ \bibinfo {author}
  {\bibfnamefont {D.}~\bibnamefont {Hwang}},\ }\bibfield  {title} {\bibinfo
  {title} {Complex networks: Structure and dynamics},\ }\href
  {https://doi.org/https://doi.org/10.1016/j.physrep.2005.10.009} {\bibfield
  {journal} {\bibinfo  {journal} {Phys. Rep.}\ }\textbf {\bibinfo {volume}
  {424}},\ \bibinfo {pages} {175} (\bibinfo {year} {2006})}\BibitemShut
  {NoStop}%
\bibitem [{\citenamefont {Barrat}\ \emph {et~al.}(2004)\citenamefont {Barrat},
  \citenamefont {Barth{\'e}lemy}, \citenamefont {Pastor-Satorras},\ and\
  \citenamefont {Vespignani}}]{Bar-04}%
  \BibitemOpen
  \bibfield  {author} {\bibinfo {author} {\bibfnamefont {A.}~\bibnamefont
  {Barrat}}, \bibinfo {author} {\bibfnamefont {M.}~\bibnamefont
  {Barth{\'e}lemy}}, \bibinfo {author} {\bibfnamefont {R.}~\bibnamefont
  {Pastor-Satorras}},\ and\ \bibinfo {author} {\bibfnamefont {A.}~\bibnamefont
  {Vespignani}},\ }\bibfield  {title} {\bibinfo {title} {The architecture of
  complex weighted networks},\ }\href {https://doi.org/10.1073/pnas.0400087101}
  {\bibfield  {journal} {\bibinfo  {journal} {Proc. Natl. Acad. Sci. U.S.A.}\
  }\textbf {\bibinfo {volume} {101}},\ \bibinfo {pages} {3747} (\bibinfo {year}
  {2004})}\BibitemShut {NoStop}%
\bibitem [{\citenamefont {Newman}(2010)}]{New-10}%
  \BibitemOpen
  \bibfield  {author} {\bibinfo {author} {\bibfnamefont {M.~E.~J.}\
  \bibnamefont {Newman}},\ }\href
  {https://doi.org/10.1093/acprof:oso/9780199206650.001.0001} {\emph {\bibinfo
  {title} {Networks: An Introduction}}}\ (\bibinfo  {publisher} {Oxford
  University Press, Oxford},\ \bibinfo {year} {2010})\BibitemShut {NoStop}%
\bibitem [{\citenamefont {Pastor-Satorras}\ \emph {et~al.}(2015)\citenamefont
  {Pastor-Satorras}, \citenamefont {Castellano}, \citenamefont {Van~Mieghem},\
  and\ \citenamefont {Vespignani}}]{Pas-15}%
  \BibitemOpen
  \bibfield  {author} {\bibinfo {author} {\bibfnamefont {R.}~\bibnamefont
  {Pastor-Satorras}}, \bibinfo {author} {\bibfnamefont {C.}~\bibnamefont
  {Castellano}}, \bibinfo {author} {\bibfnamefont {P.}~\bibnamefont
  {Van~Mieghem}},\ and\ \bibinfo {author} {\bibfnamefont {A.}~\bibnamefont
  {Vespignani}},\ }\bibfield  {title} {\bibinfo {title} {Epidemic processes in
  complex networks},\ }\href
  {https://doi.org/https://doi.org/10.1103/RevModPhys.87.925} {\bibfield
  {journal} {\bibinfo  {journal} {Rev. Mod. Phys.}\ }\textbf {\bibinfo {volume}
  {87}},\ \bibinfo {pages} {925} (\bibinfo {year} {2015})}\BibitemShut
  {NoStop}%
\bibitem [{\citenamefont {Kermack}\ and\ \citenamefont
  {McKendrick}(1927)}]{Ker-27}%
  \BibitemOpen
  \bibfield  {author} {\bibinfo {author} {\bibfnamefont {W.~O.}\ \bibnamefont
  {Kermack}}\ and\ \bibinfo {author} {\bibfnamefont {A.~G.}\ \bibnamefont
  {McKendrick}},\ }\bibfield  {title} {\bibinfo {title} {A contribution to the
  mathematical theory of epidemics},\ }\href@noop {} {\bibfield  {journal}
  {\bibinfo  {journal} {Proc. R. Soc. London A}\ }\textbf {\bibinfo {volume}
  {115}},\ \bibinfo {pages} {700} (\bibinfo {year} {1927})}\BibitemShut
  {NoStop}%
\bibitem [{\citenamefont {Grassberger}(1983)}]{Gra-83}%
  \BibitemOpen
  \bibfield  {author} {\bibinfo {author} {\bibfnamefont {P.}~\bibnamefont
  {Grassberger}},\ }\bibfield  {title} {\bibinfo {title} {On the critical
  behavior of the general epidemic process and dynamical percolation},\ }\href
  {https://doi.org/10.1016/0025-5564(82)90036-0} {\bibfield  {journal}
  {\bibinfo  {journal} {Math. Biosci.}\ }\textbf {\bibinfo {volume} {63}},\
  \bibinfo {pages} {157} (\bibinfo {year} {1983})}\BibitemShut {NoStop}%
\bibitem [{\citenamefont {Newman}(2002)}]{New-02}%
  \BibitemOpen
  \bibfield  {author} {\bibinfo {author} {\bibfnamefont {M.~E.~J.}\
  \bibnamefont {Newman}},\ }\bibfield  {title} {\bibinfo {title} {Spread of
  epidemic disease on networks},\ }\href@noop {} {\bibfield  {journal}
  {\bibinfo  {journal} {Phys. Rev. E}\ }\textbf {\bibinfo {volume} {66}},\
  \bibinfo {pages} {016128} (\bibinfo {year} {2002})}\BibitemShut {NoStop}%
\bibitem [{\citenamefont {Miller}(2007)}]{Mil-07}%
  \BibitemOpen
  \bibfield  {author} {\bibinfo {author} {\bibfnamefont {J.~C.}\ \bibnamefont
  {Miller}},\ }\bibfield  {title} {\bibinfo {title} {Epidemic size and
  probability in populations with heterogeneous infectivity and
  susceptibility},\ }\href {https://doi.org/10.1103/PhysRevE.76.010101}
  {\bibfield  {journal} {\bibinfo  {journal} {Phys. Rev. E}\ }\textbf {\bibinfo
  {volume} {76}},\ \bibinfo {pages} {010101(R)} (\bibinfo {year}
  {2007})}\BibitemShut {NoStop}%
\bibitem [{\citenamefont {Kenah}\ and\ \citenamefont {Robins}(2007)}]{Ken-07}%
  \BibitemOpen
  \bibfield  {author} {\bibinfo {author} {\bibfnamefont {E.}~\bibnamefont
  {Kenah}}\ and\ \bibinfo {author} {\bibfnamefont {J.~M.}\ \bibnamefont
  {Robins}},\ }\bibfield  {title} {\bibinfo {title} {Second look at the spread
  of epidemics on networks},\ }\href
  {https://doi.org/10.1103/PhysRevE.76.036113} {\bibfield  {journal} {\bibinfo
  {journal} {Phys. Rev. E}\ }\textbf {\bibinfo {volume} {76}},\ \bibinfo
  {pages} {036113} (\bibinfo {year} {2007})}\BibitemShut {NoStop}%
\bibitem [{\citenamefont {Lagorio}\ \emph {et~al.}(2009)\citenamefont
  {Lagorio}, \citenamefont {Migueles}, \citenamefont {Braunstein},
  \citenamefont {L\'opez},\ and\ \citenamefont {Macri}}]{Lag-09}%
  \BibitemOpen
  \bibfield  {author} {\bibinfo {author} {\bibfnamefont {C.}~\bibnamefont
  {Lagorio}}, \bibinfo {author} {\bibfnamefont {M.~V.}\ \bibnamefont
  {Migueles}}, \bibinfo {author} {\bibfnamefont {L.~A.}\ \bibnamefont
  {Braunstein}}, \bibinfo {author} {\bibfnamefont {E.}~\bibnamefont
  {L\'opez}},\ and\ \bibinfo {author} {\bibfnamefont {P.~A.}\ \bibnamefont
  {Macri}},\ }\bibfield  {title} {\bibinfo {title} {Effects of epidemic
  threshold definition on disease spread statistics},\ }\href
  {https://doi.org/10.1016/j.physa.2008.10.045} {\bibfield  {journal} {\bibinfo
   {journal} {Physica A}\ }\textbf {\bibinfo {volume} {388}},\ \bibinfo {pages}
  {755} (\bibinfo {year} {2009})}\BibitemShut {NoStop}%
\bibitem [{\citenamefont {Cohen}\ \emph {et~al.}(2003)\citenamefont {Cohen},
  \citenamefont {Havlin},\ and\ \citenamefont {{ben-Avraham}}}]{Coh-03}%
  \BibitemOpen
  \bibfield  {author} {\bibinfo {author} {\bibfnamefont {R.}~\bibnamefont
  {Cohen}}, \bibinfo {author} {\bibfnamefont {S.}~\bibnamefont {Havlin}},\ and\
  \bibinfo {author} {\bibfnamefont {D.}~\bibnamefont {{ben-Avraham}}},\
  }\bibfield  {title} {\bibinfo {title} {Efficient immunization strategies for
  computer networks and populations},\ }\href
  {https://doi.org/10.1103/PhysRevLett.91.247901} {\bibfield  {journal}
  {\bibinfo  {journal} {Phys. Rev. Lett.}\ }\textbf {\bibinfo {volume} {91}},\
  \bibinfo {pages} {247901} (\bibinfo {year} {2003})}\BibitemShut {NoStop}%
\bibitem [{\citenamefont {Ferrari}\ \emph {et~al.}(2006)\citenamefont
  {Ferrari}, \citenamefont {Bansal}, \citenamefont {Meyers},\ and\
  \citenamefont {Bj{\o}rnstad}}]{Fer-06}%
  \BibitemOpen
  \bibfield  {author} {\bibinfo {author} {\bibfnamefont {M.~J.}\ \bibnamefont
  {Ferrari}}, \bibinfo {author} {\bibfnamefont {S.}~\bibnamefont {Bansal}},
  \bibinfo {author} {\bibfnamefont {L.~A.}\ \bibnamefont {Meyers}},\ and\
  \bibinfo {author} {\bibfnamefont {O.~N.}\ \bibnamefont {Bj{\o}rnstad}},\
  }\bibfield  {title} {\bibinfo {title} {Network frailty and the geometry of
  herd immunity},\ }\href {https://doi.org/10.1098/rspb.2006.3636} {\bibfield
  {journal} {\bibinfo  {journal} {Proc. R. Soc. London, Ser. B}\ }\textbf
  {\bibinfo {volume} {273}},\ \bibinfo {pages} {2743} (\bibinfo {year}
  {2006})}\BibitemShut {NoStop}%
\bibitem [{\citenamefont {Bansal}\ \emph {et~al.}(2006)\citenamefont {Bansal},
  \citenamefont {Pourbohloul},\ and\ \citenamefont {Meyers}}]{Ban-06}%
  \BibitemOpen
  \bibfield  {author} {\bibinfo {author} {\bibfnamefont {S.}~\bibnamefont
  {Bansal}}, \bibinfo {author} {\bibfnamefont {B.}~\bibnamefont
  {Pourbohloul}},\ and\ \bibinfo {author} {\bibfnamefont {L.~A.}\ \bibnamefont
  {Meyers}},\ }\bibfield  {title} {\bibinfo {title} {A comparative analysis of
  influenza vaccination programs},\ }\href
  {https://doi.org/10.1371/journal.pmed.0030387} {\bibfield  {journal}
  {\bibinfo  {journal} {PLoS Med.}\ }\textbf {\bibinfo {volume} {3}},\ \bibinfo
  {pages} {e387} (\bibinfo {year} {2006})}\BibitemShut {NoStop}%
\bibitem [{\citenamefont {Buono}\ and\ \citenamefont
  {Braunstein}(2015)}]{Buo-15}%
  \BibitemOpen
  \bibfield  {author} {\bibinfo {author} {\bibfnamefont {C.}~\bibnamefont
  {Buono}}\ and\ \bibinfo {author} {\bibfnamefont {L.~A.}\ \bibnamefont
  {Braunstein}},\ }\bibfield  {title} {\bibinfo {title} {Immunization strategy
  for epidemic spreading on multilayer networks},\ }\href
  {https://doi.org/10.1209/0295-5075/109/26001} {\bibfield  {journal} {\bibinfo
   {journal} {Europhysics Letters}\ }\textbf {\bibinfo {volume} {109}},\
  \bibinfo {pages} {26001} (\bibinfo {year} {2015})}\BibitemShut {NoStop}%
\bibitem [{\citenamefont {Alvarez-Zuzek}\ \emph {et~al.}(2015)\citenamefont
  {Alvarez-Zuzek}, \citenamefont {Buono},\ and\ \citenamefont
  {Braunstein}}]{Zuz-15}%
  \BibitemOpen
  \bibfield  {author} {\bibinfo {author} {\bibfnamefont {L.~G.}\ \bibnamefont
  {Alvarez-Zuzek}}, \bibinfo {author} {\bibfnamefont {C.}~\bibnamefont
  {Buono}},\ and\ \bibinfo {author} {\bibfnamefont {L.~A.}\ \bibnamefont
  {Braunstein}},\ }\bibfield  {title} {\bibinfo {title} {Epidemic spreading and
  immunization strategy in multiplex networks},\ }\href
  {https://doi.org/10.1088/1742-6596/640/1/012007} {\bibfield  {journal}
  {\bibinfo  {journal} {Journal of Physics: Conference Series}\ }\textbf
  {\bibinfo {volume} {640}},\ \bibinfo {pages} {012007} (\bibinfo {year}
  {2015})}\BibitemShut {NoStop}%
\bibitem [{\citenamefont {Merler}\ \emph {et~al.}(2016)\citenamefont {Merler},
  \citenamefont {Ajelli}, \citenamefont {Fumanelli}, \citenamefont
  {Parlamento}, \citenamefont {Pastore~y Piontti}, \citenamefont {Dean},
  \citenamefont {Putoto}, \citenamefont {Carraro}, \citenamefont {Longini},
  \citenamefont {Halloran},\ and\ \citenamefont {Vespignani}}]{Mer-16}%
  \BibitemOpen
  \bibfield  {author} {\bibinfo {author} {\bibfnamefont {S.}~\bibnamefont
  {Merler}}, \bibinfo {author} {\bibfnamefont {M.}~\bibnamefont {Ajelli}},
  \bibinfo {author} {\bibfnamefont {L.}~\bibnamefont {Fumanelli}}, \bibinfo
  {author} {\bibfnamefont {S.}~\bibnamefont {Parlamento}}, \bibinfo {author}
  {\bibfnamefont {A.}~\bibnamefont {Pastore~y Piontti}}, \bibinfo {author}
  {\bibfnamefont {N.~E.}\ \bibnamefont {Dean}}, \bibinfo {author}
  {\bibfnamefont {G.}~\bibnamefont {Putoto}}, \bibinfo {author} {\bibfnamefont
  {D.}~\bibnamefont {Carraro}}, \bibinfo {author} {\bibfnamefont {I.~M.}\
  \bibnamefont {Longini}, \bibfnamefont {Jr.}}, \bibinfo {author}
  {\bibfnamefont {M.~E.}\ \bibnamefont {Halloran}},\ and\ \bibinfo {author}
  {\bibfnamefont {A.}~\bibnamefont {Vespignani}},\ }\bibfield  {title}
  {\bibinfo {title} {Containing {E}bola at the source with ring vaccination},\
  }\href {https://doi.org/10.1371/journal.pntd.0005093} {\bibfield  {journal}
  {\bibinfo  {journal} {PLOS Neglected Tropical Diseases}\ }\textbf {\bibinfo
  {volume} {10}},\ \bibinfo {pages} {1} (\bibinfo {year} {2016})}\BibitemShut
  {NoStop}%
\bibitem [{\citenamefont {Di~Muro}\ \emph {et~al.}(2018)\citenamefont
  {Di~Muro}, \citenamefont {Alvarez-Zuzek}, \citenamefont {Havlin},\ and\
  \citenamefont {Braunstein}}]{DiMu-18}%
  \BibitemOpen
  \bibfield  {author} {\bibinfo {author} {\bibfnamefont {M.~A.}\ \bibnamefont
  {Di~Muro}}, \bibinfo {author} {\bibfnamefont {L.~G.}\ \bibnamefont
  {Alvarez-Zuzek}}, \bibinfo {author} {\bibfnamefont {S.}~\bibnamefont
  {Havlin}},\ and\ \bibinfo {author} {\bibfnamefont {L.~A.}\ \bibnamefont
  {Braunstein}},\ }\bibfield  {title} {\bibinfo {title} {Multiple outbreaks in
  epidemic spreading with local vaccination and limited vaccines},\ }\href
  {https://doi.org/10.1088/1367-2630/aad723} {\bibfield  {journal} {\bibinfo
  {journal} {New J. Phys.}\ }\textbf {\bibinfo {volume} {20}},\ \bibinfo
  {pages} {083025} (\bibinfo {year} {2018})}\BibitemShut {NoStop}%
\bibitem [{\citenamefont {Alvarez-Zuzek}\ \emph {et~al.}(2019)\citenamefont
  {Alvarez-Zuzek}, \citenamefont {Di~Muro}, \citenamefont {Havlin},\ and\
  \citenamefont {Braunstein}}]{Zuz-19}%
  \BibitemOpen
  \bibfield  {author} {\bibinfo {author} {\bibfnamefont {L.~G.}\ \bibnamefont
  {Alvarez-Zuzek}}, \bibinfo {author} {\bibfnamefont {M.~A.}\ \bibnamefont
  {Di~Muro}}, \bibinfo {author} {\bibfnamefont {S.}~\bibnamefont {Havlin}},\
  and\ \bibinfo {author} {\bibfnamefont {L.~A.}\ \bibnamefont {Braunstein}},\
  }\bibfield  {title} {\bibinfo {title} {Dynamic vaccination in partially
  overlapped multiplex network},\ }\href
  {https://doi.org/10.1103/PhysRevE.99.012302} {\bibfield  {journal} {\bibinfo
  {journal} {Phys. Rev. E}\ }\textbf {\bibinfo {volume} {99}},\ \bibinfo
  {pages} {012302} (\bibinfo {year} {2019})}\BibitemShut {NoStop}%
\bibitem [{\citenamefont {Gross}\ \emph {et~al.}(2006)\citenamefont {Gross},
  \citenamefont {D'Lima},\ and\ \citenamefont {Blasius}}]{Gro-06}%
  \BibitemOpen
  \bibfield  {author} {\bibinfo {author} {\bibfnamefont {T.}~\bibnamefont
  {Gross}}, \bibinfo {author} {\bibfnamefont {C.~J.~D.}\ \bibnamefont
  {D'Lima}},\ and\ \bibinfo {author} {\bibfnamefont {B.}~\bibnamefont
  {Blasius}},\ }\bibfield  {title} {\bibinfo {title} {Epidemic dynamics on an
  adaptive network},\ }\href {https://doi.org/10.1103/PhysRevLett.96.208701}
  {\bibfield  {journal} {\bibinfo  {journal} {Phys. Rev. Lett.}\ }\textbf
  {\bibinfo {volume} {96}},\ \bibinfo {pages} {208701} (\bibinfo {year}
  {2006})}\BibitemShut {NoStop}%
\bibitem [{\citenamefont {Eastwood}\ \emph {et~al.}(2010)\citenamefont
  {Eastwood}, \citenamefont {Durrheim}, \citenamefont {Butler},\ and\
  \citenamefont {Jon}}]{Eas-10}%
  \BibitemOpen
  \bibfield  {author} {\bibinfo {author} {\bibfnamefont {K.}~\bibnamefont
  {Eastwood}}, \bibinfo {author} {\bibfnamefont {D.~N.}\ \bibnamefont
  {Durrheim}}, \bibinfo {author} {\bibfnamefont {M.}~\bibnamefont {Butler}},\
  and\ \bibinfo {author} {\bibfnamefont {E.}~\bibnamefont {Jon}},\ }\bibfield
  {title} {\bibinfo {title} {Responses to pandemic ({H1N1}) 2009,
  {A}ustralia},\ }\href {https://doi.org/10.3201/eid1608.100132} {\bibfield
  {journal} {\bibinfo  {journal} {Emerg. Infect. Dis.}\ }\textbf {\bibinfo
  {volume} {16}},\ \bibinfo {pages} {1211} (\bibinfo {year}
  {2010})}\BibitemShut {NoStop}%
\bibitem [{\citenamefont {Lagorio}\ \emph {et~al.}(2011)\citenamefont
  {Lagorio}, \citenamefont {Dickison}, \citenamefont {Vazquez}, \citenamefont
  {Braunstein}, \citenamefont {Macri}, \citenamefont {Migueles}, \citenamefont
  {Havlin},\ and\ \citenamefont {Stanley}}]{Lag-11}%
  \BibitemOpen
  \bibfield  {author} {\bibinfo {author} {\bibfnamefont {C.}~\bibnamefont
  {Lagorio}}, \bibinfo {author} {\bibfnamefont {M.}~\bibnamefont {Dickison}},
  \bibinfo {author} {\bibfnamefont {F.}~\bibnamefont {Vazquez}}, \bibinfo
  {author} {\bibfnamefont {L.~A.}\ \bibnamefont {Braunstein}}, \bibinfo
  {author} {\bibfnamefont {P.~A.}\ \bibnamefont {Macri}}, \bibinfo {author}
  {\bibfnamefont {M.~V.}\ \bibnamefont {Migueles}}, \bibinfo {author}
  {\bibfnamefont {S.}~\bibnamefont {Havlin}},\ and\ \bibinfo {author}
  {\bibfnamefont {H.~E.}\ \bibnamefont {Stanley}},\ }\bibfield  {title}
  {\bibinfo {title} {Quarantine-generated phase transition in epidemic
  spreading},\ }\href {https://doi.org/10.1103/PhysRevE.83.026102} {\bibfield
  {journal} {\bibinfo  {journal} {Phys. Rev. E}\ }\textbf {\bibinfo {volume}
  {83}},\ \bibinfo {pages} {026102} (\bibinfo {year} {2011})}\BibitemShut
  {NoStop}%
\bibitem [{\citenamefont {Buono}\ \emph {et~al.}(2012)\citenamefont {Buono},
  \citenamefont {Lagorio}, \citenamefont {Macri},\ and\ \citenamefont
  {Braunstein}}]{Buo-12}%
  \BibitemOpen
  \bibfield  {author} {\bibinfo {author} {\bibfnamefont {C.}~\bibnamefont
  {Buono}}, \bibinfo {author} {\bibfnamefont {C.}~\bibnamefont {Lagorio}},
  \bibinfo {author} {\bibfnamefont {P.~A.}\ \bibnamefont {Macri}},\ and\
  \bibinfo {author} {\bibfnamefont {L.~A.}\ \bibnamefont {Braunstein}},\
  }\bibfield  {title} {\bibinfo {title} {Crossover from weak to strong disorder
  regime in the duration of epidemics},\ }\href
  {https://doi.org/10.1016/j.physa.2012.04.002} {\bibfield  {journal} {\bibinfo
   {journal} {Physica A}\ }\textbf {\bibinfo {volume} {391}},\ \bibinfo {pages}
  {4181} (\bibinfo {year} {2012})}\BibitemShut {NoStop}%
\bibitem [{\citenamefont {Valdez}\ \emph {et~al.}(2012)\citenamefont {Valdez},
  \citenamefont {Macri},\ and\ \citenamefont {Braunstein}}]{Val-12}%
  \BibitemOpen
  \bibfield  {author} {\bibinfo {author} {\bibfnamefont {L.~D.}\ \bibnamefont
  {Valdez}}, \bibinfo {author} {\bibfnamefont {P.~A.}\ \bibnamefont {Macri}},\
  and\ \bibinfo {author} {\bibfnamefont {L.~A.}\ \bibnamefont {Braunstein}},\
  }\bibfield  {title} {\bibinfo {title} {Intermittent social distancing
  strategy for epidemic control},\ }\href@noop {} {\bibfield  {journal}
  {\bibinfo  {journal} {Phys. Rev. E}\ }\textbf {\bibinfo {volume} {85}},\
  \bibinfo {pages} {036108} (\bibinfo {year} {2012})}\BibitemShut {NoStop}%
\bibitem [{\citenamefont {Buono}\ \emph {et~al.}(2013)\citenamefont {Buono},
  \citenamefont {Vazquez}, \citenamefont {Macri},\ and\ \citenamefont
  {Braunstein}}]{Buo-13}%
  \BibitemOpen
  \bibfield  {author} {\bibinfo {author} {\bibfnamefont {C.}~\bibnamefont
  {Buono}}, \bibinfo {author} {\bibfnamefont {F.}~\bibnamefont {Vazquez}},
  \bibinfo {author} {\bibfnamefont {P.~A.}\ \bibnamefont {Macri}},\ and\
  \bibinfo {author} {\bibfnamefont {L.~A.}\ \bibnamefont {Braunstein}},\
  }\bibfield  {title} {\bibinfo {title} {Slow epidemic extinction in
  populations with heterogeneous infection rates},\ }\href
  {https://doi.org/10.1103/PhysRevE.88.022813} {\bibfield  {journal} {\bibinfo
  {journal} {Phys. Rev. E}\ }\textbf {\bibinfo {volume} {88}},\ \bibinfo
  {pages} {022813} (\bibinfo {year} {2013})}\BibitemShut {NoStop}%
\bibitem [{\citenamefont {Perez}\ \emph {et~al.}(2020)\citenamefont {Perez},
  \citenamefont {Trunfio}, \citenamefont {Rocca},\ and\ \citenamefont
  {Braunstein}}]{Per-19}%
  \BibitemOpen
  \bibfield  {author} {\bibinfo {author} {\bibfnamefont {I.~A.}\ \bibnamefont
  {Perez}}, \bibinfo {author} {\bibfnamefont {P.~A.}\ \bibnamefont {Trunfio}},
  \bibinfo {author} {\bibfnamefont {C.~E.~L.}\ \bibnamefont {Rocca}},\ and\
  \bibinfo {author} {\bibfnamefont {L.~A.}\ \bibnamefont {Braunstein}},\
  }\bibfield  {title} {\bibinfo {title} {Controlling distant contacts to reduce
  disease spreading on disordered complex networks},\ }\href
  {https://doi.org/https://doi.org/10.1016/j.physa.2019.123709} {\bibfield
  {journal} {\bibinfo  {journal} {Physica A}\ }\textbf {\bibinfo {volume}
  {545}},\ \bibinfo {pages} {123709} (\bibinfo {year} {2020})}\BibitemShut
  {NoStop}%
\bibitem [{\citenamefont {{World Health
  Organization}}(2020{\natexlab{b}})}]{WHO-20-bis}%
  \BibitemOpen
  \bibfield  {author} {\bibinfo {author} {\bibnamefont {{World Health
  Organization}}},\ }\href
  {https://www.who.int/news-room/q-a-detail/q-a-coronaviruses} {\bibinfo
  {title} {Q\&{A} on coronaviruses ({COVID}-19)}} (\bibinfo {year}
  {2020}{\natexlab{b}})\BibitemShut {NoStop}%
\bibitem [{\citenamefont {Badr}\ \emph {et~al.}(2020)\citenamefont {Badr},
  \citenamefont {Du}, \citenamefont {Marshall}, \citenamefont {Dong},
  \citenamefont {Squire},\ and\ \citenamefont {Gardner}}]{Badr-20}%
  \BibitemOpen
  \bibfield  {author} {\bibinfo {author} {\bibfnamefont {H.~S.}\ \bibnamefont
  {Badr}}, \bibinfo {author} {\bibfnamefont {H.}~\bibnamefont {Du}}, \bibinfo
  {author} {\bibfnamefont {M.}~\bibnamefont {Marshall}}, \bibinfo {author}
  {\bibfnamefont {E.}~\bibnamefont {Dong}}, \bibinfo {author} {\bibfnamefont
  {M.~M.}\ \bibnamefont {Squire}},\ and\ \bibinfo {author} {\bibfnamefont
  {L.~M.}\ \bibnamefont {Gardner}},\ }\bibfield  {title} {\bibinfo {title}
  {Association between mobility patterns and {COVID}-19 transmission in the
  {USA}: a mathematical modelling study},\ }\href@noop {} {\bibfield  {journal}
  {\bibinfo  {journal} {Lancet Infect. Dis.}\ } (\bibinfo {year}
  {2020})}\BibitemShut {NoStop}%
\bibitem [{\citenamefont {Karsai}\ \emph {et~al.}(2011)\citenamefont {Karsai},
  \citenamefont {Kivel\"a}, \citenamefont {Pan}, \citenamefont {Kaski},
  \citenamefont {Kert\'esz}, \citenamefont {Barab\'asi},\ and\ \citenamefont
  {Saram\"aki}}]{Kars-11}%
  \BibitemOpen
  \bibfield  {author} {\bibinfo {author} {\bibfnamefont {M.}~\bibnamefont
  {Karsai}}, \bibinfo {author} {\bibfnamefont {M.}~\bibnamefont {Kivel\"a}},
  \bibinfo {author} {\bibfnamefont {R.~K.}\ \bibnamefont {Pan}}, \bibinfo
  {author} {\bibfnamefont {K.}~\bibnamefont {Kaski}}, \bibinfo {author}
  {\bibfnamefont {J.}~\bibnamefont {Kert\'esz}}, \bibinfo {author}
  {\bibfnamefont {A.-L.}\ \bibnamefont {Barab\'asi}},\ and\ \bibinfo {author}
  {\bibfnamefont {J.}~\bibnamefont {Saram\"aki}},\ }\bibfield  {title}
  {\bibinfo {title} {Small but slow world: How network topology and burstiness
  slow down spreading},\ }\href {https://doi.org/10.1103/PhysRevE.83.025102}
  {\bibfield  {journal} {\bibinfo  {journal} {Phys. Rev. E}\ }\textbf {\bibinfo
  {volume} {83}},\ \bibinfo {pages} {025102(R)} (\bibinfo {year}
  {2011})}\BibitemShut {NoStop}%
\bibitem [{\citenamefont {Stehlé}\ \emph {et~al.}(2011)\citenamefont
  {Stehlé}, \citenamefont {Barrat}, \citenamefont {Cattuto}, \citenamefont
  {Pinton}, \citenamefont {Isella},\ and\ \citenamefont {den
  Broeck}}]{Steh-11}%
  \BibitemOpen
  \bibfield  {author} {\bibinfo {author} {\bibfnamefont {J.}~\bibnamefont
  {Stehlé}}, \bibinfo {author} {\bibfnamefont {A.}~\bibnamefont {Barrat}},
  \bibinfo {author} {\bibfnamefont {C.}~\bibnamefont {Cattuto}}, \bibinfo
  {author} {\bibfnamefont {J.~F.}\ \bibnamefont {Pinton}}, \bibinfo {author}
  {\bibfnamefont {L.}~\bibnamefont {Isella}},\ and\ \bibinfo {author}
  {\bibfnamefont {W.~V.}\ \bibnamefont {den Broeck}},\ }\bibfield  {title}
  {\bibinfo {title} {What’s in a crowd? {A}nalysis of face-to-face behavioral
  networks},\ }\href@noop {} {\bibfield  {journal} {\bibinfo  {journal} {J.
  Theor. Biol.}\ }\textbf {\bibinfo {volume} {271}},\ \bibinfo {pages} {166}
  (\bibinfo {year} {2011})}\BibitemShut {NoStop}%
\bibitem [{\citenamefont {Newman}(2005)}]{New-05}%
  \BibitemOpen
  \bibfield  {author} {\bibinfo {author} {\bibfnamefont {M.~E.~J.}\
  \bibnamefont {Newman}},\ }\bibfield  {title} {\bibinfo {title} {Threshold
  effects for two pathogens spreading on a network},\ }\href
  {https://doi.org/10.1103/PhysRevLett.95.108701} {\bibfield  {journal}
  {\bibinfo  {journal} {Phys. Rev. Lett.}\ }\textbf {\bibinfo {volume} {95}},\
  \bibinfo {pages} {108701} (\bibinfo {year} {2005})}\BibitemShut {NoStop}%
\bibitem [{\citenamefont {Gao}\ \emph {et~al.}(2014)\citenamefont {Gao},
  \citenamefont {Li},\ and\ \citenamefont {Havlin}}]{Jia-14}%
  \BibitemOpen
  \bibfield  {author} {\bibinfo {author} {\bibfnamefont {J.}~\bibnamefont
  {Gao}}, \bibinfo {author} {\bibfnamefont {D.}~\bibnamefont {Li}},\ and\
  \bibinfo {author} {\bibfnamefont {S.}~\bibnamefont {Havlin}},\ }\bibfield
  {title} {\bibinfo {title} {From a single network to a network of networks},\
  }\href {https://doi.org/10.1093/nsr/nwu020} {\bibfield  {journal} {\bibinfo
  {journal} {National Science Review}\ }\textbf {\bibinfo {volume} {1}},\
  \bibinfo {pages} {346} (\bibinfo {year} {2014})}\BibitemShut {NoStop}%
\bibitem [{\citenamefont {Kivelä}\ \emph {et~al.}(2014)\citenamefont
  {Kivelä}, \citenamefont {Arenas}, \citenamefont {Barthelemy}, \citenamefont
  {Gleeson}, \citenamefont {Moreno},\ and\ \citenamefont {Porter}}]{Kiv-14}%
  \BibitemOpen
  \bibfield  {author} {\bibinfo {author} {\bibfnamefont {M.}~\bibnamefont
  {Kivelä}}, \bibinfo {author} {\bibfnamefont {A.}~\bibnamefont {Arenas}},
  \bibinfo {author} {\bibfnamefont {M.}~\bibnamefont {Barthelemy}}, \bibinfo
  {author} {\bibfnamefont {J.~P.}\ \bibnamefont {Gleeson}}, \bibinfo {author}
  {\bibfnamefont {Y.}~\bibnamefont {Moreno}},\ and\ \bibinfo {author}
  {\bibfnamefont {M.~A.}\ \bibnamefont {Porter}},\ }\bibfield  {title}
  {\bibinfo {title} {Multilayer networks},\ }\href
  {https://doi.org/10.1093/comnet/cnu016} {\bibfield  {journal} {\bibinfo
  {journal} {J. Complex Netw.}\ }\textbf {\bibinfo {volume} {2}},\ \bibinfo
  {pages} {203} (\bibinfo {year} {2014})}\BibitemShut {NoStop}%
\bibitem [{\citenamefont {Boccaletti}\ \emph {et~al.}(2014)\citenamefont
  {Boccaletti}, \citenamefont {Bianconi}, \citenamefont {Criado}, \citenamefont
  {del Genio}, \citenamefont {Gómez-Gardeñes}, \citenamefont {Romance},
  \citenamefont {Sendiña-Nadal}, \citenamefont {Wang},\ and\ \citenamefont
  {Zanin}}]{Bocc-14}%
  \BibitemOpen
  \bibfield  {author} {\bibinfo {author} {\bibfnamefont {S.}~\bibnamefont
  {Boccaletti}}, \bibinfo {author} {\bibfnamefont {G.}~\bibnamefont
  {Bianconi}}, \bibinfo {author} {\bibfnamefont {R.}~\bibnamefont {Criado}},
  \bibinfo {author} {\bibfnamefont {C.}~\bibnamefont {del Genio}}, \bibinfo
  {author} {\bibfnamefont {J.}~\bibnamefont {Gómez-Gardeñes}}, \bibinfo
  {author} {\bibfnamefont {M.}~\bibnamefont {Romance}}, \bibinfo {author}
  {\bibfnamefont {I.}~\bibnamefont {Sendiña-Nadal}}, \bibinfo {author}
  {\bibfnamefont {Z.}~\bibnamefont {Wang}},\ and\ \bibinfo {author}
  {\bibfnamefont {M.}~\bibnamefont {Zanin}},\ }\bibfield  {title} {\bibinfo
  {title} {The structure and dynamics of multilayer networks},\ }\href
  {https://doi.org/https://doi.org/10.1016/j.physrep.2014.07.001} {\bibfield
  {journal} {\bibinfo  {journal} {Physics Reports}\ }\textbf {\bibinfo {volume}
  {544}},\ \bibinfo {pages} {1 } (\bibinfo {year} {2014})}\BibitemShut
  {NoStop}%
\bibitem [{\citenamefont {Kenett}\ \emph {et~al.}(2015)\citenamefont {Kenett},
  \citenamefont {Perc},\ and\ \citenamefont {Boccaletti}}]{Kene-15}%
  \BibitemOpen
  \bibfield  {author} {\bibinfo {author} {\bibfnamefont {D.~Y.}\ \bibnamefont
  {Kenett}}, \bibinfo {author} {\bibfnamefont {M.}~\bibnamefont {Perc}},\ and\
  \bibinfo {author} {\bibfnamefont {S.}~\bibnamefont {Boccaletti}},\ }\bibfield
   {title} {\bibinfo {title} {Networks of networks - {A}n introduction},\
  }\href {https://doi.org/https://doi.org/10.1016/j.chaos.2015.03.016}
  {\bibfield  {journal} {\bibinfo  {journal} {Chaos, Solitons \& Fractals}\
  }\textbf {\bibinfo {volume} {80}},\ \bibinfo {pages} {1 } (\bibinfo {year}
  {2015})}\BibitemShut {NoStop}%
\bibitem [{\citenamefont {Buldyrev}\ \emph {et~al.}(2010)\citenamefont
  {Buldyrev}, \citenamefont {Parshani}, \citenamefont {Paul}, \citenamefont
  {Stanley},\ and\ \citenamefont {Havlin}}]{Bul-10}%
  \BibitemOpen
  \bibfield  {author} {\bibinfo {author} {\bibfnamefont {S.~V.}\ \bibnamefont
  {Buldyrev}}, \bibinfo {author} {\bibfnamefont {R.}~\bibnamefont {Parshani}},
  \bibinfo {author} {\bibfnamefont {G.}~\bibnamefont {Paul}}, \bibinfo {author}
  {\bibfnamefont {H.~E.}\ \bibnamefont {Stanley}},\ and\ \bibinfo {author}
  {\bibfnamefont {S.}~\bibnamefont {Havlin}},\ }\bibfield  {title} {\bibinfo
  {title} {{Catastrophic cascade of failures in interdependent networks}},\
  }\href {https://doi.org/10.1038/nature08932} {\bibfield  {journal} {\bibinfo
  {journal} {Nature (London)}\ }\textbf {\bibinfo {volume} {464}},\ \bibinfo
  {pages} {1025} (\bibinfo {year} {2010})}\BibitemShut {NoStop}%
\bibitem [{\citenamefont {Brummitt}\ \emph {et~al.}(2012)\citenamefont
  {Brummitt}, \citenamefont {D{\textquoteright}Souza},\ and\ \citenamefont
  {Leicht}}]{Bru-12}%
  \BibitemOpen
  \bibfield  {author} {\bibinfo {author} {\bibfnamefont {C.~D.}\ \bibnamefont
  {Brummitt}}, \bibinfo {author} {\bibfnamefont {R.~M.}\ \bibnamefont
  {D{\textquoteright}Souza}},\ and\ \bibinfo {author} {\bibfnamefont {E.~A.}\
  \bibnamefont {Leicht}},\ }\bibfield  {title} {\bibinfo {title} {Suppressing
  cascades of load in interdependent networks},\ }\href
  {https://doi.org/10.1073/pnas.1110586109} {\bibfield  {journal} {\bibinfo
  {journal} {Proc. Natl. Acad. of Sci USA}\ }\textbf {\bibinfo {volume}
  {109}},\ \bibinfo {pages} {E680} (\bibinfo {year} {2012})}\BibitemShut
  {NoStop}%
\bibitem [{\citenamefont {Di~Muro}\ \emph {et~al.}(2016)\citenamefont
  {Di~Muro}, \citenamefont {Buldyrev}, \citenamefont {Stanley},\ and\
  \citenamefont {Braunstein}}]{DiMu-16}%
  \BibitemOpen
  \bibfield  {author} {\bibinfo {author} {\bibfnamefont {M.~A.}\ \bibnamefont
  {Di~Muro}}, \bibinfo {author} {\bibfnamefont {S.~V.}\ \bibnamefont
  {Buldyrev}}, \bibinfo {author} {\bibfnamefont {H.~E.}\ \bibnamefont
  {Stanley}},\ and\ \bibinfo {author} {\bibfnamefont {L.~A.}\ \bibnamefont
  {Braunstein}},\ }\bibfield  {title} {\bibinfo {title} {Cascading failures in
  interdependent networks with finite functional components},\ }\href@noop {}
  {\bibfield  {journal} {\bibinfo  {journal} {Phys. Rev. E}\ }\textbf {\bibinfo
  {volume} {94}},\ \bibinfo {pages} {042304} (\bibinfo {year}
  {2016})}\BibitemShut {NoStop}%
\bibitem [{\citenamefont {Castellano}\ \emph {et~al.}(2009)\citenamefont
  {Castellano}, \citenamefont {Fortunato},\ and\ \citenamefont
  {Loreto}}]{Cast-09}%
  \BibitemOpen
  \bibfield  {author} {\bibinfo {author} {\bibfnamefont {C.}~\bibnamefont
  {Castellano}}, \bibinfo {author} {\bibfnamefont {S.}~\bibnamefont
  {Fortunato}},\ and\ \bibinfo {author} {\bibfnamefont {V.}~\bibnamefont
  {Loreto}},\ }\bibfield  {title} {\bibinfo {title} {{Statistical physics of
  social dynamics}},\ }\href {https://doi.org/10.1103/RevModPhys.81.591}
  {\bibfield  {journal} {\bibinfo  {journal} {Rev. Mod. Phys.}\ }\textbf
  {\bibinfo {volume} {81}},\ \bibinfo {pages} {591} (\bibinfo {year}
  {2009})}\BibitemShut {NoStop}%
\bibitem [{\citenamefont {Galam}(2008)}]{Gal-08}%
  \BibitemOpen
  \bibfield  {author} {\bibinfo {author} {\bibfnamefont {S.}~\bibnamefont
  {Galam}},\ }\bibfield  {title} {\bibinfo {title} {Sociophysics: A review of
  {G}alam models},\ }\href {https://doi.org/10.1142/S0129183108012297}
  {\bibfield  {journal} {\bibinfo  {journal} {International Journal of Modern
  Physics C}\ }\textbf {\bibinfo {volume} {19}},\ \bibinfo {pages} {409}
  (\bibinfo {year} {2008})}\BibitemShut {NoStop}%
\bibitem [{\citenamefont {Saumell-Mendiola}\ \emph {et~al.}(2012)\citenamefont
  {Saumell-Mendiola}, \citenamefont {Serrano},\ and\ \citenamefont
  {Bogu{\~n}{\'a}}}]{Sau-12}%
  \BibitemOpen
  \bibfield  {author} {\bibinfo {author} {\bibfnamefont {A.}~\bibnamefont
  {Saumell-Mendiola}}, \bibinfo {author} {\bibfnamefont {M.~{\'A}.}\
  \bibnamefont {Serrano}},\ and\ \bibinfo {author} {\bibfnamefont
  {M.}~\bibnamefont {Bogu{\~n}{\'a}}},\ }\bibfield  {title} {\bibinfo {title}
  {{Epidemic spreading on interconnected networks.}},\ }\href
  {https://doi.org/10.1103/PhysRevE.86.026106} {\bibfield  {journal} {\bibinfo
  {journal} {Phys. Rev. E}\ }\textbf {\bibinfo {volume} {86}},\ \bibinfo
  {pages} {026106} (\bibinfo {year} {2012})}\BibitemShut {NoStop}%
\bibitem [{\citenamefont {Cozzo}\ \emph {et~al.}(2013)\citenamefont {Cozzo},
  \citenamefont {Ba{\~n}os}, \citenamefont {Meloni},\ and\ \citenamefont
  {Moreno}}]{Coz-13}%
  \BibitemOpen
  \bibfield  {author} {\bibinfo {author} {\bibfnamefont {E.}~\bibnamefont
  {Cozzo}}, \bibinfo {author} {\bibfnamefont {R.~A.}\ \bibnamefont
  {Ba{\~n}os}}, \bibinfo {author} {\bibfnamefont {S.}~\bibnamefont {Meloni}},\
  and\ \bibinfo {author} {\bibfnamefont {Y.}~\bibnamefont {Moreno}},\
  }\bibfield  {title} {\bibinfo {title} {Contact-based social contagion in
  multiplex networks},\ }\href {https://doi.org/10.1103/PhysRevE.88.050801}
  {\bibfield  {journal} {\bibinfo  {journal} {Phys. Rev. E}\ }\textbf {\bibinfo
  {volume} {88}},\ \bibinfo {pages} {050801(R)} (\bibinfo {year}
  {2013})}\BibitemShut {NoStop}%
\bibitem [{\citenamefont {De~Domenico}\ \emph {et~al.}(2016)\citenamefont
  {De~Domenico}, \citenamefont {Granell}, \citenamefont {Porter},\ and\
  \citenamefont {Arenas}}]{Are-16}%
  \BibitemOpen
  \bibfield  {author} {\bibinfo {author} {\bibfnamefont {M.}~\bibnamefont
  {De~Domenico}}, \bibinfo {author} {\bibfnamefont {C.}~\bibnamefont
  {Granell}}, \bibinfo {author} {\bibfnamefont {M.~A.}\ \bibnamefont
  {Porter}},\ and\ \bibinfo {author} {\bibfnamefont {A.}~\bibnamefont
  {Arenas}},\ }\bibfield  {title} {\bibinfo {title} {The physics of spreading
  processes in multilayer networks},\ }\href@noop {} {\bibfield  {journal}
  {\bibinfo  {journal} {Nature Physics}\ }\textbf {\bibinfo {volume} {12}},\
  \bibinfo {pages} {901} (\bibinfo {year} {2016})}\BibitemShut {NoStop}%
\bibitem [{\citenamefont {Buono}\ \emph {et~al.}(2014)\citenamefont {Buono},
  \citenamefont {Alvarez-Zuzek}, \citenamefont {Macri},\ and\ \citenamefont
  {Braunstein}}]{Buo-14}%
  \BibitemOpen
  \bibfield  {author} {\bibinfo {author} {\bibfnamefont {C.}~\bibnamefont
  {Buono}}, \bibinfo {author} {\bibfnamefont {L.~G.}\ \bibnamefont
  {Alvarez-Zuzek}}, \bibinfo {author} {\bibfnamefont {P.~A.}\ \bibnamefont
  {Macri}},\ and\ \bibinfo {author} {\bibfnamefont {L.~A.}\ \bibnamefont
  {Braunstein}},\ }\bibfield  {title} {\bibinfo {title} {Epidemics in partially
  overlapped multiplex networks},\ }\href
  {https://doi.org/10.1371/journal.pone.0092200} {\bibfield  {journal}
  {\bibinfo  {journal} {PLoS ONE}\ }\textbf {\bibinfo {volume} {9}},\ \bibinfo
  {pages} {e92200} (\bibinfo {year} {2014})}\BibitemShut {NoStop}%
\bibitem [{\citenamefont {Molloy}\ and\ \citenamefont {Reed}(1995)}]{Moll-95}%
  \BibitemOpen
  \bibfield  {author} {\bibinfo {author} {\bibfnamefont {M.}~\bibnamefont
  {Molloy}}\ and\ \bibinfo {author} {\bibfnamefont {B.}~\bibnamefont {Reed}},\
  }\bibfield  {title} {\bibinfo {title} {A critical point for random graphs
  with a given degree sequence},\ }\href
  {https://doi.org/10.1002/rsa.3240060204} {\bibfield  {journal} {\bibinfo
  {journal} {Random Struct. Algor.}\ }\textbf {\bibinfo {volume} {6}},\
  \bibinfo {pages} {161} (\bibinfo {year} {1995})}\BibitemShut {NoStop}%
\bibitem [{\citenamefont {Erd\H{o}s}\ and\ \citenamefont
  {R\'enyi}(1959)}]{Erd-59}%
  \BibitemOpen
  \bibfield  {author} {\bibinfo {author} {\bibfnamefont {P.}~\bibnamefont
  {Erd\H{o}s}}\ and\ \bibinfo {author} {\bibfnamefont {A.}~\bibnamefont
  {R\'enyi}},\ }\bibfield  {title} {\bibinfo {title} {On random graphs {I}},\
  }\href@noop {} {\bibfield  {journal} {\bibinfo  {journal} {Publicationes
  Mathematicae Debrecen}\ }\textbf {\bibinfo {volume} {6}},\ \bibinfo {pages}
  {290} (\bibinfo {year} {1959})}\BibitemShut {NoStop}%
\bibitem [{\citenamefont {Braunstein}\ \emph {et~al.}(2007)\citenamefont
  {Braunstein}, \citenamefont {Wu}, \citenamefont {Chen}, \citenamefont
  {Buldyrev}, \citenamefont {Kalisy}, \citenamefont {Sreenivasan},
  \citenamefont {Cohen}, \citenamefont {L\'opez}, \citenamefont {Havlin},\ and\
  \citenamefont {Stanley}}]{Brau-07}%
  \BibitemOpen
  \bibfield  {author} {\bibinfo {author} {\bibfnamefont {L.~A.}\ \bibnamefont
  {Braunstein}}, \bibinfo {author} {\bibfnamefont {Z.}~\bibnamefont {Wu}},
  \bibinfo {author} {\bibfnamefont {Y.}~\bibnamefont {Chen}}, \bibinfo {author}
  {\bibfnamefont {S.~V.}\ \bibnamefont {Buldyrev}}, \bibinfo {author}
  {\bibfnamefont {T.}~\bibnamefont {Kalisy}}, \bibinfo {author} {\bibfnamefont
  {S.}~\bibnamefont {Sreenivasan}}, \bibinfo {author} {\bibfnamefont
  {R.}~\bibnamefont {Cohen}}, \bibinfo {author} {\bibfnamefont
  {E.}~\bibnamefont {L\'opez}}, \bibinfo {author} {\bibfnamefont
  {S.}~\bibnamefont {Havlin}},\ and\ \bibinfo {author} {\bibfnamefont {H.~E.}\
  \bibnamefont {Stanley}},\ }\bibfield  {title} {\bibinfo {title} {Optimal path
  and minimal spanning trees in random weighted networks},\ }\href
  {https://doi.org/10.1142/S0218127407018361} {\bibfield  {journal} {\bibinfo
  {journal} {International Journal of Bifurcation and Chaos}\ }\textbf
  {\bibinfo {volume} {17}},\ \bibinfo {pages} {2215} (\bibinfo {year}
  {2007})}\BibitemShut {NoStop}%
\bibitem [{\citenamefont {Valdez}\ \emph {et~al.}(2013)\citenamefont {Valdez},
  \citenamefont {Buono}, \citenamefont {Macri},\ and\ \citenamefont
  {Braunstein}}]{Val-13}%
  \BibitemOpen
  \bibfield  {author} {\bibinfo {author} {\bibfnamefont {L.~D.}\ \bibnamefont
  {Valdez}}, \bibinfo {author} {\bibfnamefont {C.}~\bibnamefont {Buono}},
  \bibinfo {author} {\bibfnamefont {P.~A.}\ \bibnamefont {Macri}},\ and\
  \bibinfo {author} {\bibfnamefont {L.~A.}\ \bibnamefont {Braunstein}},\
  }\bibfield  {title} {\bibinfo {title} {{Social distancing strategies against
  disease spreading}},\ }\href@noop {} {\bibfield  {journal} {\bibinfo
  {journal} {Fractals}\ }\textbf {\bibinfo {volume} {21}},\ \bibinfo {pages}
  {1350019} (\bibinfo {year} {2013})}\BibitemShut {NoStop}%
\bibitem [{\citenamefont {Newman}\ \emph {et~al.}(2001)\citenamefont {Newman},
  \citenamefont {Strogatz},\ and\ \citenamefont {Watts}}]{New-01}%
  \BibitemOpen
  \bibfield  {author} {\bibinfo {author} {\bibfnamefont {M.~E.~J.}\
  \bibnamefont {Newman}}, \bibinfo {author} {\bibfnamefont {S.~H.}\
  \bibnamefont {Strogatz}},\ and\ \bibinfo {author} {\bibfnamefont {D.~J.}\
  \bibnamefont {Watts}},\ }\bibfield  {title} {\bibinfo {title} {Random graphs
  with arbitrary degree distributions and their applications},\ }\href@noop {}
  {\bibfield  {journal} {\bibinfo  {journal} {Phys. Rev. E}\ }\textbf {\bibinfo
  {volume} {64}},\ \bibinfo {pages} {026118} (\bibinfo {year}
  {2001})}\BibitemShut {NoStop}%
\bibitem [{\citenamefont {Newman}(2003)}]{New-03}%
  \BibitemOpen
  \bibfield  {author} {\bibinfo {author} {\bibfnamefont {M.~E.~J.}\
  \bibnamefont {Newman}},\ }\bibfield  {title} {\bibinfo {title} {The structure
  and function of complex networks},\ }\href
  {https://doi.org/10.1137/S003614450342480} {\bibfield  {journal} {\bibinfo
  {journal} {SIAM Rev.}\ }\textbf {\bibinfo {volume} {45}},\ \bibinfo {pages}
  {167} (\bibinfo {year} {2003})}\BibitemShut {NoStop}%
\bibitem [{\citenamefont {Wang}\ \emph {et~al.}(2017)\citenamefont {Wang},
  \citenamefont {Tang}, \citenamefont {Stanley},\ and\ \citenamefont
  {Braunstein}}]{Brau-17}%
  \BibitemOpen
  \bibfield  {author} {\bibinfo {author} {\bibfnamefont {W.}~\bibnamefont
  {Wang}}, \bibinfo {author} {\bibfnamefont {M.}~\bibnamefont {Tang}}, \bibinfo
  {author} {\bibfnamefont {H.~E.}\ \bibnamefont {Stanley}},\ and\ \bibinfo
  {author} {\bibfnamefont {L.~A.}\ \bibnamefont {Braunstein}},\ }\bibfield
  {title} {\bibinfo {title} {Unification of theoretical approaches for epidemic
  spreading on complex networks},\ }\href@noop {} {\bibfield  {journal}
  {\bibinfo  {journal} {Reports on Progress in Physics}\ }\textbf {\bibinfo
  {volume} {80}},\ \bibinfo {pages} {036603} (\bibinfo {year}
  {2017})}\BibitemShut {NoStop}%
\end{thebibliography}
\end{document}